\documentclass[aps,prb,twocolumn,groupedaddress]{revtex4-1}\usepackage{hyperref}
\usepackage{amsmath}
\usepackage{amssymb}
\usepackage{bm}
\usepackage{bbold}
\usepackage{graphicx}

\usepackage{hyperref}
\hypersetup{%
  breaklinks = {true},
  citecolor = {blue},
  colorlinks = {true},
  linkcolor = {red},
  pdfauthor = {\textcopyright\ Matteo Bazzanella},
  pdfcreator = {\LaTeX\ and \flqq hyperref\frqq},
  pdffitwindow = {true},
  pdfmenubar = {true},
  pdfpagelayout = {SinglePage},
  pdfstartview = {Fit},
  pdftoolbar = {true},
  plainpages = {false},
}

\newcommand{\ua}{\uparrow}
\newcommand{\da}{\downarrow}
\newcommand{\dd}{\dagger}
\newcommand{\be}{\begin{eqnarray}}
\newcommand{\ee}{\end{eqnarray}}
\newenvironment{sistema}%
{\left\lbrace\begin{array}{@{}l@{}}}%
{\end{array}\right.}

\usepackage{color}
\usepackage{color}
\usepackage{color}

\begin{document}

\title{Non-Linear Methods in Strongly Correlated Electron Systems}

\author{Matteo Bazzanella and Johan Nilsson}
\address{Department of Physics, University of Gothenburg, 412 96 Gothenburg,  Sweden}
%\ead{johan.nilsson@physics.gu.se}

\date{August 2014}

\begin{abstract}
We analyze the structure of the group of (local) non-linear canonical transformations that exist in a system with $n$ fermionic modes.
To perform our study we develop an alternative framework to represent the generators of these canonical transformations;
indeed we show how their definition, understanding and control is significantly improved using the Majorana fermion representation.
These canonical transformations have the structure of a Lie group and we provide a
representation for the elements of the Lie algebra that is very convenient both conceptually and practically (computationally):
indeed our framework yields as side product an extremely effective tool to handle and work with $SU(2^n)$ Lie groups and algebras.
Granting an enhanced control of the group of non-linear canonical transformation, our framework can be helpful in the study of strongly correlated electron systems,
since it allows to easily identify fermionic degrees of freedom able to capture part of the correlations and thus may give a simpler representation of the Hamiltonian.
Thanks to our analysis, also symmetry-based studies of the quantum Hamiltonians can be improved, since the simple representation of the generators
of the canonical group permits to identify and understand otherwise hidden symmetries of difficult interpretation.
The main aim of this work is to provide a comprehensive, general and scrupulous analysis of this framework,
that we already applied in some circumstances. Therefore specific applications will not be presented in detail, since they can be found elsewhere,
while only the formal and conceptual aspects will be developed thoroughly.
%We already applied our scheme in the analysis of physically relevant models and we will only review here the main results.
%Moreover, we outline relevant scenarios where the applications of our approach may increase our predictive power, dedicating
%special attention to the correlated hopping models.
\end{abstract}

\maketitle

\section{Introduction}
The absence of a universally accepted strategy for the study of strongly correlated electron systems (SCES) is one of the major
obstacles for physicists attempting their study. Although many numerical and analytical techniques
have been developed to perform the analysis of an increasing number of systems and model Hamiltonians, it is still
very difficult to identify a reliable and effective method that can be applied in any circumstance. This is due to the fact that even the
most modern numerical techniques are seriously challenged by the unusual physics of SCES.
Many recent debates in the SCES community, for example in the context of cuprates \cite{Phillips:2010aa},
are making more and more evident the limitations of analysis based on the original electron degrees of freedom and Landau's theory,
advocating the need of a formalism that is more flexible in the definition of the fundamental degrees of freedom.
The development of a framework that posses such a versatility is the subject of this work.
In particular our study simplifies the use of non-linear canonical transformations in the analysis of the fermionic quantum problems.
The idea that non-linear transformation can be useful in this context is not completely new\cite{Harris:1967qf,StellanMele,StellanKONDO}.
Such transformations allow to study the SCES from a different perspective, since the original fermionic degrees of freedom play no central role,
permitting the identification of customized quantum coordinates to describe the correlated systems.
Our analysis is based on the Majorana fermion representation of the fermion and spin operators.
Making use of this representation it becomes possible to easily identify all the possible local canonical transformations and to understand the structure of their transformation group.
We focus our attention on the fact that these transformations have the well defined mathematical structure of a Lie group (beside a $\mathbb{Z}_2$ discrete component),
which we fully characterize in the general case of $n$ fermion species, identifying the generators of all its continuous components and determining their
underlying algebraic structure.
Once that a convenient form for the generators is found, it becomes extremely easy to control all the non-linear canonical transformations;
moreover, in this representation, it becomes also simple to understand the origin of the new (transformed) degrees of freedom.
In terms of the transformed fermions (or spins) the lattice model Hamiltonians gets often simplified, making them more accessible to the available methods.
In particular, the identification of the generators of the groups of non-linear transformations in terms of Majorana fermions
permits also the analysis of the Hamiltonians in terms of symmetries that may otherwise be difficult to find using the standard representation in terms of fermionic operators.

%\news{(skip and move into maintext) Our discussion is clearly not the most general possible, since it relies on the concept of non-linear canonical transformation and naturally favors
%fermionic degrees of freedom. More comprehensive approaches\cite{Scharnhorst:2011aa}, which try to mix bosonic and fermionic
%degrees of freedom, have been studied for finite small systems, but these methods will not be part of our analysis.}

We already obtained interesting results working with the framework that we suggest\cite{JohanKondoORIGINAL,ChristmaMajorana,Bazzanella:2014aa,MBzNilssonPREP}.
We will briefly review them here for completeness, but we encourage the reader to analyze our other works, in order to have
concrete examples of the effectiveness of our framework. In this sense, the present paper must be understood as the theoretical counterpart of the others,
which were focused on the application of the framework, but were lacking of mathematical insight, analysis and necessary generalization.

Beside the straightforward application to the description of non-linear canonical transformations, our results provide also a very convenient (economic) representation
of the elements of the Lie groups $SU(2^n)$ and of the generators of their Lie algebra, in terms of Majorana fermions.
This in turns simplifies the application, also in the SCES context, of some concepts such as spectrum generating algebras and dynamical symmetries, familiar in nuclear and particle physics.
Also this possibility offered by the Majorana fermion representation should be taken in consideration in future applications of our framework.

This paper is organized as follows:
in Sec.~\ref{sec:section1} we review the concept of non-linear canonical transformation of the quantum degrees of freedom,
we determine the structure of such a transformation group, we develop the representation that permits to handle it in an easy way and we
characterize its generators in the most general case of $n$ fermion species. This section contains clearly the major novelties introduced in this paper.
In Sec.~\ref{sec:applications} we will review some situations where our formalism is a powerful tool, which have been subject of other works.
In Sec.~\ref{sec:noncan} we will explain how these concepts are useful in the definition of the holon-hyperspin representation,
which offer an interesting insight on some quantum problems.

\section{Non-linear canonical transformations}\label{sec:section1}
\subsection{Example: the four-dimensional Hilbert space}\label{sec:sec1example}
Let us consider a system composed of an arbitrary $N\ge1$ number of sites and consider one of the $N$ local Hilbert spaces, assuming that
the system contains only one single spinful fermion species.
This is the situation encountered for example in the Hubbard model, where the local Hilbert spaces $\mathcal{H_H}$
are spanned by the four basis states $\left\{|\Omega\rangle,|\ua\rangle,|\da\rangle,|\ua\da\rangle\right\}$,
which are obtained by applying two fermionic operators $c^\dd_\da$ and $c^\dd_\ua$ on a properly defined local vacuum state $|\Omega\rangle$.
%So by definition:
%\be
%&&\exists\quad |\Omega\rangle\quad | \quad c_\da|\Omega\rangle=c_\ua|\Omega\rangle=0, \nonumber\\
%&&\text{with}\quad\{c_\da,c_\ua\}=\{c^\dagger_\da,c^\dagger_\ua\}=0,\nonumber\\
%&&\text{and}\quad\{c_\da,c^\dagger_\da\}=\{c_\ua,c^\dagger_\ua\}=1.\nonumber
%\ee
The presence of the fermionic operators $c^\dd_\da$, $c^\dd_\ua$ (and hermitian conjugates) implies
the existence of a Fock structure on the Hilbert space. This structure is clearly not unique, but it depends upon the specific definition
of the operators $c^\dd_\da$, $c^\dd_\ua$, and hence upon the choice of the four basis states of the local Hilbert space.
We will define the {\it canonical-group} as
the group of transformations that is able to change the Fock structure of a certain Hilbert space, i.e., the group that starting from an original
set of fermionic operators ($c_\da$, $c_\ua$ in this case) is able to generate a new set of fermionic operators ($d_\da$, $d_\ua$)
acting on the same Hilbert space.
If $\mathcal{H_H}$ was the entire Hilbert space, then the canonical group would coincide\cite{StellanMele} with the group of transformations that preserves
\be\label{eq:globalcanonical}\begin{split}
&\{c^\dd_\sigma,c_{\sigma^\prime}\}=\delta_{\sigma\sigma^\prime},&\\
&\{c_\sigma,c_{\sigma^\prime}\}=0,&
\end{split}
\ee
which is $SU(4)$. But, since $\mathcal{H_H}$ is only the local Hilbert space associated with one single site $r$,
the previous relation is not the only one that should be satisfied; instead the more general constraint should be fulfilled:
\be\label{eq:StellanConstraint}\begin{split}
&\{c^\dd_\sigma (r),c_{\sigma^\prime}(r^\prime)\}=\delta_{\sigma\sigma^\prime}\delta_{rr^\prime},&\\
&\{c_\sigma (r),c_{\sigma^\prime}(r^\prime)\}=0.&
\end{split}
\ee
This reduces the group to
\be\label{eq:canonicalgrHH}
SU(2)\otimes SU(2)\otimes U(1) \otimes \mathbb{Z}_2.
\ee
In the rest of the manuscript the term {\it (Local) Canonical Group} (LCG) will always refer to the group of transformations that fulfill (\ref{eq:StellanConstraint}), when applied on each lattice site.
Therefore we will only deal with local transformations and coordinates, allowing us to neglect (in most cases) the lattice index of the operators, for sake of notation.
A discussion of these transformations can be found in Ref.~\onlinecite{StellanMele} or in Appendix \ref{app:Wenger}, where we review and generalize these concepts.
We stress the fact that in this manuscript we discuss transformations of the {\it representation} of the quantum degrees of freedom,
in other words the transformations among the different equivalent ways that we can choose to discuss the physics of a given system.

%\news{If... allora il gruppo combacierebbe con quello che preserva $\{c^\dd,cc\}$ ovvero SU(4). Ma siccome lattice allora e' un sottogruppo... etc....}
%This group must be contained in the group $SU(4)$, where the factor four comes from the dimension of $\mathcal{H_H}$. In fact, the canonical-group
%generates a new orthogonormal basis set for the Hilbert space. Since the latter is a (finite) complex vector space, the most general transformation
%that can realize this result is $SU(4)$. \news{(Why not anti-unitary transf? Are the discrete parts really contained?)}

%The canonical-group of $\mathcal{H_H}$ has been characterized long ago\cite{StellanMele},
%and it has been identified with
%$$
%SU(2)\otimes SU(2)\otimes U(1) \otimes \mathbb{Z}_2.
%$$
The group (\ref{eq:canonicalgrHH}) is composed of three parts: two linear transformation subgroups $SU(2)$,
a discrete part $\mathbb{Z}_2$ and a subgroup $U(1)$ that contains the non-linear transformations.
The linear/non-linear nature of the transformation depends upon the type of combination
of the original fermionic operators performed by it.
The linear transformations $SU(2)$ are generated by the spin $S_i$ and charge-isospin $I_i$ operators; instead the
discrete set $\mathbb{Z}_2$ contains the particle-hole exchange transformation for only one fermion species.
This latter discrete part does not change the basis states of the Hilbert space, but it affects only the Fock structure, changing how the fermions are counted.
The most interesting part of the canonical-group is given by the non-linear set $U(1)$, which generates transformations like
\be\label{StellansNL}
\begin{split}
c^\dagger_\ua &\rightarrow& {c^\prime}^\dd_\ua=c^\dagger_\ua(1-c^\dagger_\da c_\da)+e^{2i \chi}c^\dagger_\ua c^\dagger_\da c_\da,\label{eq:NLtransf}\\
c^\dagger_\da &\rightarrow& {c^\prime}^\dd_\da=c^\dagger_\da(1-c^\dagger_\ua c_\ua)+e^{2i \chi}c^\dagger_\da c^\dagger_\ua c_\ua,
\end{split}
\ee
defining the correlated fermions (operators) ${c^\prime}^\dd_\ua$ and ${c^\prime}^\dd_\da$.
%All (?>) these transformations turn out to be (local) symmetries of the Hubbard model for some choice of the parameters and their systematic study is
%convenient in order to understand the phase diagram of the model.
%Although the existence of this canonical group, in the case of the Hubbard model, has been known for some time, no one ever succeeded in a systematic
%characterization of its structure in higher dimensional Fock spaces \news{(I HOPE)}. Therefore
%no systematic identification of all the symmetries of any second quantized Hamiltonian has been carried out.
%In the following we will show what is the general structure of the canonical-group for an
%arbitrary big Hilbert space. Thanks to the identification of this structure it will be possible to quantitatively improve the
%analysis of quantum systems that are based on fermionic degrees of freedom. In particular it will change the way how we look
%at the quantum systems, making more general (and flexible) our idea of fermions and setting up new quantum numbers
%more appropriate for the description of the quantum levels (hyperspin quantum numbers).
While it is very easy to understand the action of the linear part of the group, it is more difficult to understand how and why the
non-linear transformation exists. The Majorana representation plays a crucial role in allowing insight in this case.

As a first step we must switch from the fermionic representation of the quantum degrees
of freedom, to a different one given in terms of Majorana fermions. To do this, we observe that we can always think of
a fermionic operator $c_n$ as $c_n=(\gamma_{2n-1}-i\gamma_{2n})/2$, if the operators $\gamma_{2n-1},\gamma_{2n}$ are properly defined.
These operators are called Majorana fermions (Majoranas). In general, given a set of $n$ fermions, then $2n$ Majoranas are needed.
To form proper fermionic operators, generating the anti-commutation relations correctly,
the Majoranas have to close to Clifford algebra, so that
\be\label{Clifforddef}
\left\{\gamma_i,\gamma_j\right\}=2\delta_{ij},\quad
\gamma_i^\dagger=\gamma_i,\quad \gamma_i^2=1.
\ee
Of course the argument can be reversed: given a set of $2n$ operators fulfilling the
relations (\ref{Clifforddef}) and acting on a $2^n$ dimensional Hilbert space, then $n$ fermionic operators can be defined combining them linearly.
Such fermionic operators are able to span the entire Hilbert space (which will therefore also have a specific Fock structure).

%During the years, many Hamiltonian have been studied making use of this Majorana decomposition.
%What has been overlooked is the fact that the Clifford structure of the Majoranas algebra generates
%a much bigger set of Majorana fermions, that can be used to represent more general (but perfectly well defined) fermionic operators.
In the rest of the manuscript we will use the following convention: to form a fermion (operator) the Majoranas must be combined in the following way
\be\label{convention}
d_m=\frac{\pm\gamma_{2m-1}- i\gamma_{2m}}{2},\quad m\in\{1,..,n\},
\ee
where the (conventional) role of $\pm$ sign will be clarified later on.
This is not the most general definition, but all the other definitions can be connected to this one via a transformation
that belongs to the canonical-group, as we will show.
The inclusion of the non-linear transformations, among the possible unitary transformations analyzed, permits the consideration of
a larger set of Majoranas and therefore a larger set of fermionic operators.

Returning to the example of $\mathcal{H_H}$, let us write down the Clifford algebra\cite{Lachieze-Rey:2010qp} generated by the Majoranas.
By definition
\be\label{eq:conventionHubbard}
c_\ua=\frac{\gamma_1-i\gamma_2}{2} \text{\quad and\quad} c_\da=\frac{-\gamma_3-i\gamma_4}{2},
\ee
where the choice of the minus signs in front of the Majorana $\gamma_3$ is purely conventional and fixed by historical reasons.
The basic elements of the Clifford algebra are:
\be
&1&\label{Hubbard:zero}\\
&\gamma_1,\quad \gamma_2, \quad \gamma_3,\quad\gamma_4,&\label{Hubbard:first}\\
&\gamma_1\gamma_2,\quad \gamma_1\gamma_3,\quad \gamma_1\gamma_4,\quad \gamma_2\gamma_3,\quad \gamma_2\gamma_4,\quad \gamma_3\gamma_4,&\label{Hubbard:second}\\
&\gamma_1\gamma_2\gamma_3,\quad \gamma_1\gamma_2\gamma_4,\quad \gamma_1\gamma_3\gamma_4,\quad \gamma_2\gamma_3\gamma_4,&\label{Hubbard:third}\\
&\gamma_1\gamma_2\gamma_3\gamma_4.\label{Hubbard:fourth}&
\ee
So according to the definition (\ref{eq:conventionHubbard}), the fermionic operators that can span the Hilbert space of the Hubbard model
are built combining, on the row (\ref{Hubbard:first}), the first element with the second one and the third element with the fourth one.

Once the two fermionic operators have been decomposed, a new couple of fermionic operators
can be defined making a (symmetric) linear combination of two Majorana operators of the first line (\ref{Hubbard:first}). For example we can define
\be\label{ex:hubbard1}
d_1=\frac{\gamma_1-i\gamma_3}{2} \text{\quad and\quad} d_2=\frac{\gamma_2-i\gamma_4}{2};
\ee
this is a (trivial) transformation of the Clifford algebra of the Majorana fermions. In practice we have changed the
order of the Majoranas on line (\ref{Hubbard:first}) and inverted the sign of the Majorana $\gamma_2$,
with respect to the definition (\ref{eq:conventionHubbard}). Indeed all these interpretations rely on the fact that we have fixed some
convention (\ref{eq:conventionHubbard}) for the formation of the fermionic operators.
Such an order exchange is performed by a transformation of the Clifford algebra, which in the specific case
is accomplished by the rotor\cite{Lachieze-Rey:2010qp} $\exp(-\frac{\theta}{2}\gamma_2\gamma_3),$ fixing $\theta=\pi/2$.
By definition, the action of a rotor on the Clifford generators ($\gamma_1,...,\gamma_4$) of the algebra is
\be\label{originalrotorexample}
\alpha_i=e^{-\frac{\theta}{2}\gamma_2\gamma_3} \gamma_i e^{\frac{\theta}{2}\gamma_2\gamma_3}.
\ee
Remembering that $(\gamma_i\gamma_j)^2=-1$ it is evident that $e^{-\frac{\theta}{2}\gamma_2\gamma_3}=\cos(\theta/2)-\gamma_2\gamma_3\sin(\theta/2)$.
%In fact
%&&e^{-\frac{\theta}{2}\gamma_2\gamma_3}=\\
%&&=1+\left(-\frac{\theta}{2}\gamma_2\gamma_3\right)+\frac{1}{2!}\left(-\frac{\theta}{2}\gamma_2\gamma_3\right)^2+\frac{1}{3!}\left(-\frac{\theta}{2}\gamma_2\gamma_3\right)^3+..\nonumber\\
%&&=1+\left(-\frac{\theta}{2}\gamma_2\gamma_3\right)-\frac{1}{2!}\left(-\frac{\theta}{2}\right)^2-\frac{1}{3!}\left(-\frac{\theta}{2}\right)^3 \gamma_2\gamma_3+...\nonumber\\
%&&=\cos(\theta/2)-\gamma_2\gamma_3\sin(\theta/2)\nonumber.
%\ee
This means that the new set of Majoranas generated by this transformation is
\be
&&\alpha_1=\gamma_1,\nonumber\\
&&\alpha_2=\cos(\theta)\gamma_2+\sin(\theta)\gamma_3,\\
&&\alpha_3=-\sin(\theta)\gamma_2+\cos(\theta)\gamma_3,\nonumber\\
&&\alpha_4=\gamma_4.\nonumber
\ee
The reader may check by direct inspection that the set $\{\alpha_i\}$ closes to Clifford algebra, otherwise it can be proved in the following way:
\be
\{\alpha_i,\alpha_j\}&=&
e^{-\frac{\theta}{2}\gamma_p\gamma_q}\gamma_i e^{\frac{\theta}{2}\gamma_p\gamma_q}e^{-\frac{\theta}{2}\gamma_p\gamma_q}\gamma_j e^{\frac{\theta}{2}\gamma_p\gamma_q}+\nonumber\\
&&\quad+ e^{-\frac{\theta}{2}\gamma_p\gamma_q}\gamma_j e^{\frac{\theta}{2}\gamma_p\gamma_q}e^{-\frac{\theta}{2}\gamma_p\gamma_q}\gamma_j e^{\frac{\theta}{2}\gamma_p\gamma_q}=\nonumber\\
&=&e^{-\frac{\theta}{2}\gamma_p\gamma_q}\gamma_i \gamma_j e^{\frac{\theta}{2}\gamma_p\gamma_q}+
e^{-\frac{\theta}{2}\gamma_p\gamma_q}\gamma_j \gamma_j e^{\frac{\theta}{2}\gamma_p\gamma_q}=\nonumber\\
&=&e^{-\frac{\theta}{2}\gamma_p\gamma_q}\{\gamma_i, \gamma_j\} e^{\frac{\theta}{2}\gamma_p\gamma_q}=\nonumber\\
&=&2\delta_{ij}\qquad i,j,p,q\in\{1,2,3,4\}.
\ee
In the same way it can be checked that the transformation of the full Clifford algebra is {\it consistent}, i.e.
\be\label{consistency}
&&\alpha_a\alpha_b=e^{-\frac{\theta}{2}\gamma_p\gamma_q} \gamma_a\gamma_b e^{\frac{\theta}{2}\gamma_p\gamma_q},\nonumber\\
&&\alpha_a\alpha_b\alpha_c=e^{-\frac{\theta}{2}\gamma_p\gamma_q} \gamma_a\gamma_b\gamma_c e^{\frac{\theta}{2}\gamma_p\gamma_q},\\
&&\alpha_1\alpha_2\alpha_3\alpha_4=e^{-\frac{\theta}{2}\gamma_p\gamma_q} \gamma_1\gamma_2\gamma_3\gamma_4 e^{\frac{\theta}{2}\gamma_p\gamma_q}=\gamma_1\gamma_2\gamma_3\gamma_4,\nonumber
\ee
with $a,b,c,p,q \in\{1,2,3,4\}$.
Since $\{\alpha_i\}$ closes to Clifford algebra the $\alpha_i$ can be used to build fermionic operators $d_1,d_2$
that will span the Hilbert space and define its Fock structure. Applying the definition (\ref{eq:conventionHubbard}):
\be
d_1&=&\frac{\alpha_1-i\alpha_2}{2}=\frac{\gamma_1-i\left\{\cos(\theta)\gamma_2+\sin(\theta)\gamma_3\right\}}{2},\\
d_2&=&\frac{-\alpha_3-i\alpha_4}{2}=\frac{-\left\{-\sin(\theta)\gamma_2+\cos(\theta)\gamma_3\right\}-i\gamma_4}{2},\nonumber
\ee
and choosing $\theta=\pi/2$ we generate the result (\ref{ex:hubbard1}):
%\be\label{ex:hubbard}
%&d_1=\frac{\alpha_1-i\alpha_2}{2}=\frac{\gamma_1-i\gamma_3}{2},& \nonumber\\
%&\text{\quad and\quad}&\\
%&d_2=\frac{-\alpha_3-i\alpha_4}{2}=\frac{\gamma_2-i\gamma_4}{2}.&\nonumber
%\ee}
\be\label{ex:hubbard}
d_1=\frac{\gamma_1-i\gamma_3}{2},\quad d_2=\frac{\gamma_2-i\gamma_4}{2}.
\ee

%In particular it can also be checked that
%\be
%\alpha_a\alpha_b=e^{-\frac{\theta}{2}\gamma_p\gamma_q} \gamma_a\gamma_b e^{\frac{\theta}{2}\gamma_p\gamma_q}=\gamma_a\gamma_b,
%\ee
%which means that these kind of transformations do not change the quadratic operators (associated with physical properties as spin and isospin).

It is useful to understand the meaning of these transformations in terms of the original fermionic operators.
The transformation realized by the rotor of our example is a simple
linear combination of the original operators $c_\da$, $c_\ua$ (and h.c.).
Indeed, the new fermions $d_1$ and $d_2$ are obtained applying
a linear transformation that is far from unconventional. In fact we can think of it as:
\be
d_1=e^{-\frac{\theta}{2}\gamma_2\gamma_3}c_\ua e^{\frac{\theta}{2}\gamma_2\gamma_3},\quad d_2=e^{-\frac{\theta}{2}\gamma_2\gamma_3}c_\da e^{\frac{\theta}{2}\gamma_2\gamma_3},
\ee
but since
\be\label{eq:spin-isospin}
S_x=-i\frac{\gamma_2\gamma_3+\gamma_1\gamma_4}{4},\quad I_x=-i\frac{\gamma_2\gamma_3-\gamma_1\gamma_4}{4},
\ee
one discovers that the transformation given by the rotor $\exp\left(-\frac{\theta}{2}\gamma_2\gamma_3\right)$ is simply $\exp\left(-i\theta(S_x+I_x)\right)$,
which is one of the linear transformations contained in $SU(2)\otimes SU(2)$. Of course the analysis holds for the other bilinears
(and linear combinations of bilinears) of line (\ref{Hubbard:second}) too. So, analyzing the full Majorana Clifford algebra, we discover
the origin of the linear part of the canonical-group of $\mathcal{H_H}$, as the group of the transformations generated by the bilinear rotors of the Clifford algebra.
These rotors transform the original Clifford algebra, mapping the original set of Majoranas into another one. This causes a change of the entire Clifford algebra,
combining the higher order terms in a consistent way.
En passant we note that the second line (\ref{Hubbard:second}) gets mixed also by the linear transformations,
and therefore new spin and isospin operators associated with the $d_i$-fermions get defined as linear combinations of the old ones.
%Hence it is still possible to use a notation $d_1=d_{\tilde\ua}$ and $d_2=d_{\tilde\da}$,
Keeping in mind 
%the arbitrariness involved in the definition of spin and isospin. In general
this arbitrariness, we will make use of the term {\it hyperspin} to indicate these quantum numbers.
The hyperspin will be defined more rigorously later on, in the text.
%This arbitrariness is responsible for the irrelevancy of the conventional $\pm$ choice in (\ref{convention}) and (\ref{eq:conventionHubbard}).
%We would like to remind to the reader that so far we did not discuss any particular Hamiltonian.
%We have seen that the spin and isospin generate canonical transformations in $\mathcal{H_H}$. Such transformations are
%represented by the elements of the subgroup $SU(2)\otimes SU(2)$ contained into $SU(2)\otimes SU(2)\otimes U(1)\otimes \mathbb{Z}_2$, and correspond to the
%transformations of the Majorana Clifford algebra generated by the rotors of the bilinear operators on the line (\ref{Hubbard:second}).

Now that the origin of the $SU(2)\otimes SU(2)$ components of the LCG of $\mathcal{H_H}$ has been understood and interpreted as transformations of the Clifford algebra,
we now move our focus on the non-linear component $U(1)$. To understand its origin one should recall a very well known
property of the Majorana fermions, i.e., the fact that by multiplying together an odd number of
Majoranas one obtains again objects that behave as Majoranas\cite{WilczekMajoranaErice}. In our example this means that
the objects of the line (\ref{Hubbard:third}), if properly multiplied by an imaginary unit,
form a new set of Majorana fermions like $\{i\gamma_i\gamma_j\gamma_k\}$. These three-composite objects can therefore be used to build
well defined fermionic operators.
One could guess that a general canonical transformation could involve (always using $\mathcal{H_H}$ as example)
a ``rotation'' between the lines (\ref{Hubbard:first}) and (\ref{Hubbard:third}) of the Majorana Clifford algebra.
In fact, mixing {\it properly} the two lines one can obtain a combination of the two sets of single and three-composite Majoranas, producing
a set of objects that still behave as Majoranas (\ref{Clifforddef}).
In this example we use the term {\it Hodge rotation} to indicate the rotation of the two lines.
It is not difficult to realize that this ``appropriate'' combination is generated by
the operator $\exp\left(-i\frac{\theta}{2}\gamma_1\gamma_2\gamma_3\gamma_4\right)$, so that
\be\label{eq:HodgeN2}
\mu_1&=&e^{-i\frac{\theta}{2}\gamma_1\gamma_2\gamma_3\gamma_4} \gamma_1 e^{i\frac{\theta}{2}\gamma_1\gamma_2\gamma_3\gamma_4}\nonumber\\
&=&\cos(\theta)\gamma_1+\sin(\theta)i\gamma_2\gamma_3\gamma_4.
\ee
One can easily check that the new set $\{\mu_i\}$ is a set of properly defined Majoranas, which close to Clifford algebra consistently,
as defined in (\ref{consistency}).
The cornerstone of this transformation is the imaginary unit in front of the operator $\gamma_1\gamma_2\gamma_3\gamma_4$. On the one hand it makes
$i\gamma_1\gamma_2\gamma_3\gamma_4$ square to $-1$, so that the algebra used in the case of the bilinear rotor is still valid and makes
the transformation unitary; on the other hand
it turns the combination of the two lines into the combination of the two sets of Majoranas mentioned previously.
It can be noticed that the bilinears $\mu_i\mu_j$ are untouched by the Hodge rotation, so $\mu_i\mu_j=\gamma_i\gamma_j$
(differently from the transformations analyzed previously).
The Hodge rotation realizes the non-linear transformation (\ref{StellansNL}). In fact
\be
\begin{split}
&\mu_1=\cos(\theta)\gamma_1+\sin(\theta)i\gamma_2\gamma_3\gamma_4,&\\
&\mu_2=\cos(\theta)\gamma_2-\sin(\theta)i\gamma_1\gamma_3\gamma_4,&\\
&\mu_3=\cos(\theta)\gamma_3+\sin(\theta)i\gamma_1\gamma_2\gamma_4,&\\
&\mu_4=\cos(\theta)\gamma_4-\sin(\theta)i\gamma_1\gamma_2\gamma_3,&
\end{split}
\ee
which means, applying again (\ref{eq:conventionHubbard}) to define the fermionic operators,
%\begin{widetext}
\be
d^\dagger_1&=&\frac{\mu_1+i\mu_2}{2}\nonumber\\
&=&\frac{\cos(\theta)(\gamma_1+i\gamma_2)+\sin(\theta)\left(i\gamma_2\gamma_3\gamma_4+\gamma_1\gamma_3\gamma_4\right)}{2}\nonumber\\
%&=&\frac{\cos(\theta)(\gamma_1)+\sin(\theta)i\gamma_2\gamma_3\gamma_4+i\left(\cos(\theta)\gamma_2-\sin(\theta)i\gamma_1\gamma_3\gamma_4\right)}{2}\nonumber\\
&=&e^{i\theta}c^\dagger_\ua-\left(e^{i\theta}-e^{-i\theta}\right)c^\dagger_\ua c^\dagger_\da c_\da,\nonumber\\
d^\dagger_2&=&e^{i\theta}c^\dd_\da-\left(e^{i\theta}-e^{-i\theta}\right)c^\dd_\da c^\dd_\ua c_\ua.
\ee
%\end{widetext}
As a consequence (\ref{eq:NLtransf}) is recovered putting $\theta=-\chi$ and removing an irrelevant phase pre-factor $e^{i\theta}$, {\it common} to $d^\dagger_1$ and $d^\dagger_2$.

It is now evident that all continuous components of the LCG of  $\mathcal{H_H}$ %(or, more precisely, of the Fock structure of $\mathcal{H_H}$)
are simply generated by transformations of the Clifford algebra of the constituents Majoranas, which are realized by the even rows of the algebra itself,
if properly multiplied by adequate imaginary units.
%It is very important to notice that no other continuous canonical-transformation can exist,
%since $SU(2)\otimes SU(2)\otimes U(1)$ is a maximal Lie subgroup of $SU(4)$. Therefore no other well defined sets of Majoranas can be built
%starting from the original Clifford algebra.
However, this does not exclude the presence of other discrete components in LCG.
These transformations cannot make linear combinations nor exchanges of Majoranas, since we have seen that these kind of operations involve continuous transformations.
In the canonical-group of $\mathcal{H_H}$, there exists only one discrete subgroup of such transformations, given by $\mathbb{Z}_2$. Such a transformation
does not make any linear combination of the basis states of the Hilbert space, but it changes the way they are labeled and how the fermions are counted.
In fact the effect of $\mathbb{Z}_2$ is
\be
c^\dagger_\ua\rightarrow c^\dagger_\ua,\quad c^\dagger_\da\rightarrow c_\da.
\ee
In terms of Majoranas this means that the sign of an odd number of the elements of the line (\ref{Hubbard:first}) changes.
A closer study shows that the only independent transformation is the one that exchanges the sign of just one Majorana, i.e., the sign
of the object obtained multiplying together all the Majoranas $\gamma_1\gamma_2\gamma_3\gamma_4$, which is proportional to the local parity of the system.
All the other possible discrete transformations can be obtained as a combination of this single exchange and one of the continuous elements of LCG.
It is clear that such sign change is actually irrelevant, since it can be reabsorbed into the conventional definition
of the fermion creation/annihilation operators~(\ref{convention}).
So, although the application of the transformation can indeed simplify the problem
(as in the case of the Shiba transformation in the analysis\cite{ShibaOriginal} of  the negative $U$ Hubbard model),
these discrete transformations are quite irrelevant from the point of view of the Majorana Clifford algebra.
%Anyway it becomes easy to count and classify them using the full Clifford algebra. In particular it is interesting to see
%how some will affect the total parity (reversing its sign) while some others will not. We will make some more specific and general examples in the next sections.

%In order to do it, we will need a better comprehension of the structure of the canonical-group of the Hilbert space.
%I'll keep the discussion the most generic possible, but at the end I will have to switch to the case of 4 (Hubbard), 6 (Kondo) and 8 (Anderson) Majoranas. However I think
%I can fix this details and prove by induction that for a generic number $2n$ of Majoranas the structure that I find holds.
%I will not use as basic example the 4 dimensional Hilbert space (Hubbard), because characterized by a too poor structure.
%So I'm forced to use the Hilbert space with 8 dimensions (Kondo).

\subsection{General case}\label{sec:section2math}

The main point that we want to make in this paper is that
it is possible to generalize all the previous concepts to larger local Hilbert spaces, which means to systems that have higher number of fermionic species.
To show how this is can be done, it is necessary to prove some statements about the group of the canonical transformations of the local Hilbert space.
%\news{In the following sections the results of this analysis are shown while the demonstrations are reported in the appendices.
%Later in the manuscript we will show how it is possible to apply this framework for the study of the SCES.}

Let us consider a local Hilbert space $\mathcal{H}$ of total dimension $2^n$, where $n$ fermion species are defined.
As we show in Appendix~\ref{app:Wenger}, the general structure of the continuous part of LCG is:
$$
SU(2^{n-1})\otimes SU(2^{n-1})\otimes U(1).
$$
We saw that with $n=2$, the rationale behind the structure of all the possible canonical transformations could be
understood by analyzing the transformations of the Clifford algebra of the Majorana
fermions associated with the fermionic operators; which also means by identifying a set of transformations that
is able to mix properly the odd elements of the Clifford algebra. This is true also for $n>2$, if a set of convenient
generators for all the continuous canonical transformations in LCG is correctly identified.
%Studying these transformations we will demonstrate that the canonical group of such a Fock space is
%$SU(2^{n-1})\otimes SU(2^{n-1})\otimes U(1)$, times a number of discrete symmetries. We will show how to build the generators of the different Lie subalgebras
%and how to characterize the relations between them.
%Given a Fock space of dimension $2^n$, a set of $2n$ Majoranas that close properly to Clifford algebra is automatically defined, 
%By construction\cite{Lachieze-Rey:2010qp} this algebra is isomorphic to $Cl(\mathbf R^{2n,0})$.

We begin taking all the elements of the full Clifford algebra generated by $2n$ Majoranas, but the trivial element:
\be\label{originalelements}
\begin{split}
&\gamma_1,\gamma_2, ... , \gamma_1\gamma_2\gamma_3,... = \Delta_i\quad \text{odd},&\\
&\gamma_1\gamma_2,...,\gamma_1\gamma_2\gamma_3\gamma_4,... = \Omega_i\quad \text{even}.&
\end{split}
\ee
For future convenience we defined the sets of even elements $\{\Omega_i\}^{(n)}$ and odd elements $\{\Delta_i\}^{(n)}$.
Our convention is that $\gamma_{i_1}\gamma_{i_2}...\gamma_{i_m}$ has $i_1<i_2<...<i_m$, with $1\leq m\leq 2n$.
To facilitate the use of this convention we will use the following notation to indicate any element obtained
multiplying together $m$ Majoranas:
\be
\epsilon_{i_1,i_2,...,i_m}\gamma_{i_1}\gamma_{i_2}...\gamma_{i_m},
\ee
with {\it no} Einstein convention and with $\epsilon_{i_1,i_2,...,i_m}$ the m-dimensional Levi-Civita symbol\footnote{Typically the Levi-Civita symbol
is defined as a totally antisymmetric object with $l$ indices that can take $l$ different values. Formally: $\epsilon_{i_1...i_{m=l}}=\Pi_{1\leq a<b\leq l} sgn(i_b-i_a)$.
In our case the indices are $m\leq l=2n$, but this clearly does not invalidate the previous definition of the symbol.}.
With this notation the specific order
of the Majoranas becomes irrelevant, since the Levi-Civita symbol in front of the multi-Majorana object returns the correct sign.

Multiplying the objects inside the sets (\ref{originalelements}) by appropriate imaginary units, we can turn them into {\it antihermitian} operators:
\be\label{randd}
\begin{split}
&i\gamma_1,i\gamma_2, ... , \gamma_1\gamma_2\gamma_3,... = d_i\quad \text{odd},&\\
&\gamma_1\gamma_2,...,i\gamma_1\gamma_2\gamma_3\gamma_4,... = p_i\quad \text{even}.&
\end{split}
\ee
The reader should note that this means $d_i^2=p_i^2=-1$.
To simplify the convention we will adopt the following notation
for the elements $p_j$ of $\{p_k\}^{(n)}$:
\be\label{eq:notazioneI}
p_j\in \{p_k\}^{(n)} \Leftrightarrow p_j=\mathcal{I}(o_j)\epsilon_{i_1,i_2,...,i_{2o_j}}\gamma_{i_1}\gamma_{i_2}...\gamma_{i_{2o_j}},
\ee
where
\be
\mathcal{I}(o_j)=+1\text{ if }o_j\text{ is odd};\quad
\mathcal{I}(o_j)=+i\text{ if }o_j\text{ is even}\nonumber.
\ee
We name with $\mathcal{T}^{(n)}$ the total set 
\be\label{defT}
\mathcal{T}^{(n)}=\{t_i\}^{(n)}=\{p_i\}^{(n)}\cup\{d_i\}^{(n)};
\ee
A fundamental result is obtained in Appendix \ref{qpp:Johansapproach}, where we prove that this set of
antihermitian operators closes to Lie algebra:
\be
\mathcal{T}^{(n)}\simeq su(2^n),
\ee
with the Lie product defined as
\be
[t_i,t_j]=t_it_j-t_jt_i,
\ee
which is appropriate since the Majoranas admit matrix representation.
Moreover we show that taking the set $\{p_i\}^{(n)}$ and {\it removing all the operators that contain one arbitrarily chosen
Majorana}, one obtains another Lie algebra $\mathcal{L}^{(n)}_{1/2}$, such that:
\be
\mathcal{L}^{(n)}_{1/2}\simeq su(2^{n-1}).
\ee
The elements in $\mathcal{T}^{(n)}$ and $\mathcal{L}^{(n)}_{1/2}$ can therefore be used to generate the groups $SU(2^n)$
and $SU(2^{n-1})$ via exponentiation (see Table \ref{tabsummary} for summary).
For reasons that will become clear later we name $\mathcal{L}^{(n)}_{1/2}$ the {\it hyperspin} algebra.

We must strongly remark the non-triviality of the two latter results, which go beyond the known results.
The most known relation between Clifford algebras and Lie groups is probably the connection between the orthogonal transformation $O(2n)$ group
and the transformations generated by all the bilinears of a Clifford algebra.
We greatly enlarge this notion showing how all the elements of the algebra can be used to build a well known Lie algebra, if properly redefined.
It is appropriate to point out how this relation may be of interest in quite a number of circumstances beyond the context of non-linear canonical transformations,
in the light of the renovated interest of the community in emergent (composite) Majorana modes\cite{WilczekMajoranaErice,Lee:2013aa},
non-abelian quantum computation\cite{Nayak:1996ms,Alicea:2011eb} and multi-wire Majorana junctions\cite{Beri:2012cq,Beri:2013aa,Zazunov:2014aa,Zazunov:2013aa}.

\begin{table*}[t]
\centering
\caption{The schematic summary of the relationship between the different Lie algebras defined in the manuscript. We recall the definitions: $\mathcal{T}^{(j)}$
is given (\ref{defT},\ref{defT2}) by the set of antihermitian operators containing all the elements of the Clifford algerba generated by $2j$ Majoranas, except the scalar element.
The set $\{p_i\}^{(j)}$ (\ref{randd}) contains instead only the elements of $\mathcal{T}^{(j)}$ that are obtained as the multiplication of an even number of Majoranas. The set $\mathcal{L}^{(j)}$
is obtained (\ref{defL}) from $\{p_i\}^{(j)}$ removing $p_{max}$, while $\mathcal{L}^{(j)}_{1/2}$
is built taking the elements of $\mathcal{L}^{(j)}$ that do not contain an arbitrarily chosen Majorana. We chose the symbol $u(1)$ to indicate
the presence of a generator that commutes with all the other and that generates a $U(1)$ Lie subgroup.}\label{tabsummary}
\begin{ruledtabular}
\begin{tabular}{rccr}
Number of Majornas & \qquad Set of antihermitian operators & Algebra & \qquad Group defined via exponentiation\\
\hline
$2m$ & $\mathcal{T}^{(m)}$ & $su(2^m)$ &$SU(2^m)$\\
	& $\{p_i\}^{(m)}=\mathcal{L}^{(m)}\oplus \{p_{max}\}$ & $su(2^{m-1})\oplus su(2^{m-1})\oplus u(1)$ & $SU(2^{m-1})\otimes SU(2^{m-1})\otimes U(1)$\\
	&  & &\\
$2m+1=2n-1$ & $\{\tilde p_i\}^{(m)}=\mathcal{L}^{(m+1)}_{1/2}=\mathcal{L}^{(n)}_{1/2}$ & $su(2^m)=su(2^{n-1})$ &$SU(2^m)=SU(2^{n-1})$\\
&&&\\
$2n=2m+2$ & $\mathcal{T}^{(n)}$ & $su(2^n)$ &$SU(2^n)$\\
	& $\{p_i\}^{(n)}=\mathcal{L}^{(n)}\oplus \{p_{max}\}$ & $su(2^{n-1})\oplus su(2^{n-1})\oplus u(1)$ & $SU(2^{n-1})\otimes SU(2^{n-1})\otimes U(1)$\\
... & ... & ... & ...\\
\end{tabular}
\end{ruledtabular}
\end{table*}

One can note that $\mathcal{L}^{(n)}_{1/2}$ does not include the top dimensional form $p_{max}=\mathcal{I}(2n)\gamma_1\gamma_2 ... \gamma_{2n}$.
Hence, such operator can be used to define two orthogonal projectotion operators $(1\pm i p_{max})/2$. In fact:
\be
\begin{split}
&\left(\frac{1\pm i p_{max}}{2}\right)^2=\frac{1\pm i p_{max}}{2},&\\
&\frac{1+ i p_{max}}{2} \frac{1- i p_{max}}{2}=0.&
\end{split}
\ee
We can define two sets $\mathcal{L}^{(n)}_\alpha$ and $\mathcal{L}^{(n)}_\beta$ multiplying all the elements
in $\mathcal{L}^{(n)}_{1/2}$ by $(1+ i p_{max})/2$ and $(1- i p_{max})/2$ respectively.
Evidently the same two sets may be obtained starting from
\be\label{defL}
\mathcal{L}^{(n)}=\{p_k\}^{(n)}\quad\text{without}\quad p_{max},
\ee
and combining properly the objects inside it.

The elements in $\mathcal{L}^{(n)}_\alpha$ and $\mathcal{L}^{(n)}_\beta$ can be written as:
\be\label{alfabeta}
\alpha_i=\frac{p_i+ip_{max}p_i}{2} ,\quad \beta_i=\frac{p_i-ip_{max}p_i}{2},
\ee
for all $p_i\in \mathcal{L}^{(n)}_{1/2}$ or equivalently $\mathcal{L}^{(n)}$;
the reader can note that by construction $ip_{max}p_i \in \mathcal{L}^{(n)},\forall p_i$. Writing the elements in this way it becomes clear that
the operators in $\mathcal{L}^{(n)}_\alpha$ and $\mathcal{L}^{(n)}_\beta$ commute with each other, since
\be
[p_i+ip_{max}p_i&,&p_j-ip_{max}p_j]=\\
&=&[p_i,p_j]+[ip_{max}p_i,p_j]+\nonumber\\
&&\quad-[p_i,ip_{max}p_j]-[ip_{max}p_i,ip_{max}p_j]\nonumber\\
%&=&[p_i,p_j]+ip_{max}[p_i,p_j]+\nonumber\\
%&&\quad-ip_{max}[p_i,p_j]-[p_i,p_j]\nonumber\\
&=&0.\nonumber
\ee
Given the properties of $\mathcal{L}^{(n)}_{1/2}$, it is easy to prove that also $\mathcal{L}^{(n)}_\alpha$ and $\mathcal{L}^{(n)}_\beta$ close to Lie algebra. In fact,
consider any couple of operators $p_i,p_j \in \mathcal{L}^{(n)}_{1/2}$ and indicate their Lie product as
\be
[p_i,p_j]=c^q_{ij} p_q,
\ee
where the structure constants $c^q_{ij}$ are computed explicitly in Appendix~\ref{app:NLTLIE}.
Take now two elements $\alpha_i,\alpha_j \in \mathcal{L}^{(n)}_\alpha$ associated with $p_i,p_j$ as
\be
\alpha_i = p_i \frac{1+ i p_{max}}{2},\quad \alpha_j = p_j \frac{1+ i p_{max}}{2},
\ee
Then
\be
[\alpha_i,\alpha_j]&=&\frac{1}{4}[(1+ip_{max})p_i,(1+ip_{max})p_j]\nonumber\\
&=&\frac{(1+ip_{max})}{2}[p_i,p_j]\nonumber\\
&=&\frac{(1+ip_{max})}{2}c^q_{ij}p_q\nonumber\\
&=&c^q_{ij}\alpha_q,
\ee
and similarly for $\beta_i,\beta_j \in \mathcal{L}^{(n)}_\beta$ and 
\be
[\beta_i,\beta_j]&=&c^q_{ij}\beta_q\nonumber,
\ee
Since the structure constants are the same, we have that
\be
\mathcal{L}^{(n)}_\beta\simeq \mathcal{L}^{(n)}_\alpha \simeq \mathcal{L}^{(n)}_{1/2}\simeq su(2^{n-1}).
\ee

Given the previous properties it is easy to understand that
\be
\mathcal{L}^{(n)}=\mathcal{L}^{(n)}_\alpha\oplus\mathcal{L}^{(n)}_\beta,
\ee
so $\mathcal{L}^{(n)}$ is a semi-simple Lie algebra and its exponentiation generates the Lie group
\be
SU(2^{n-1})\otimes SU(2^{n-1}).
\ee
In the $n=2$ case, $\mathcal{L}_\alpha^{(2)}$ and $\mathcal{L}_\beta^{(2)}$ correspond to isospin and spin algebras respectively.
This is due to the fact that the operators $(1+ ip_{max})/2$ and $(1- ip_{max})/2$ are projectors on the even and odd sectors of the Hilbert space.
Consequently we name $\mathcal{L}_\alpha^{(n)}$ the algebra of {\it isospin-sector of the hyperspin} (ISH) and $\mathcal{L}_\beta^{(n)}$ the {\it spin-sector of the hyperspin} (SSH) respectively.
These concepts will be made clear later on, but here it is important to stress that, because of these properties, no transformation generated by $\mathcal{L}^{(n)}$ can mix the even
and odd sectors of the Hilbert space.
%One should note that, in general, since the two sets $\mathcal{L}^{(n)}_\alpha$ and $\mathcal{L}^{(n)}_\beta$ commute, it is {\it always possible}
%to use the original generators defined by $\mathcal{L}^{(n)}$, using the Baker-Campbell-Hausdorff formula:
%\be
%e^{-\frac{\theta}{2}\alpha_i}e^{-\frac{\theta}{2}\beta_i}=e^{-\frac{\theta}{2}(\alpha_i+\beta_i)}=e^{-\frac{\theta}{2}p_i}.
%\ee
%This can be of great help if one wants to redefine the standard spin-isospin operators, as in (\ref{eq:spin-isospin}) for example,
%in terms of these ISH and SSH operators.
This is also true for the $U(1)$ transformations generated by $p_{max}$, since this operator is (beside prefactors) the local parity operator $P_L$, mentioned in Appendix~\ref{app:Wenger}.
Considering also this $U(1)$ group of transformations, we can understand that $\{p_i\}^{(n)}$ is a Lie algebra that generates the non-semisimple Lie group 
\be
NLT=SU(2^{n-1})\otimes SU(2^{n-1})\otimes U(1),
\ee
where $U(1)$ is an invariant subgroup and each transformation in NLT does not mix the even and odd sectors of the Hilbert space.
Since this is a maximal compact non-semisimple Lie subgroup\cite{Feger:2012fk} of $SU(2^n)$, it cannot be enlarged further by any continuos subgroup of $SU(2^n)$.

The previous results mean that NLT may be the continuos part of the local canonical group LCG of a $2^n$ dimensional Hilbert space.
To prove this point {\it one should show that NLT contains only canonical transformations}. To do it
we define the action of an element $G(\theta) \in$ NLT on the $2n$ Majorana fermions as
\be\label{def:NLT}
\mu_j=e^{-\frac{\theta_i}{2}p_i}\gamma_j e^{\frac{\theta_i}{2}p_i}\quad\forall i,j,\text{ and } \theta_i\in[0,2\pi),
\ee
consistently with (\ref{eq:HodgeN2}) and with the definition of the action of a rotor on the basic generators of the Clifford algebra\cite{Lachieze-Rey:2010qp}.
The calculation is indeed very much simplified by the fact that $p_i^2=-1$, since $e^{\frac{\theta_i}{2}p_i}=\cos\left(\theta/2\right)+p_i\sin\left(\theta/2\right)$,
so it becomes easy to check that (\ref{def:NLT}) produces a linear combination of the original Majorana $\gamma_j$,
with a candidate Majorana fermion built using an odd number of the original Majoranas, as in (\ref{eq:HodgeN2}).

Evidently the new set of objects $\mu_j$ square to $1$, are hermitian and close to Clifford algebra, in fact:
\be
\{\mu_i,\mu_j\}&=&e^{-\frac{\theta_i}{2} p_{i}}\gamma_i e^{\frac{\theta_i}{2} p_{i}}e^{-\frac{\theta_i}{2} p_{i}}\gamma_j e^{\frac{\theta_i}{2} p_{i}}+\nonumber\\
&&\quad+e^{-\frac{\theta_i}{2} p_{i}}\gamma_j e^{\frac{\theta_i}{2} p_{i}}e^{-\frac{\theta_i}{2} p_{_i}}\gamma_i e^{\frac{\theta_i}{2} p_{i}} \nonumber\\
&=&e^{-\frac{\theta_i}{2} p_{i}}\{\gamma_i \gamma_j+\gamma_j \gamma_i \}e^{\frac{\theta_i}{2} p_{i}} \nonumber\\
&=&e^{-\frac{\theta_i}{2} p_{i}}2\delta_{ij}e^{\frac{\theta_i}{2} p_{i}}=2\delta_{ij}. \nonumber
\ee
In the same way it can be shown that the transformation is {\it consistent} as defined in Sec.~\ref{sec:sec1example};
since the demonstration follows exactly the arguments in (\ref{consistency}), we will skip it.
This means that the $\mu_j$ form a well defined set of $2n$ Majorana fermions, which can be used to form new $n$ fermionic operators.
So NLT contains only canonical transformations, defined via (\ref{def:NLT}).

For example, we can consider the effect of the $U(1)$ subgroup, i.e., the transformations generated by $p_{max}$,
which acts on the set $\{\gamma_i\}$ as defined in~(\ref{def:NLT}):
\be\label{def:actiontransf1}
e^{-\frac{\theta}{2}p_{max}} &\gamma_j& e^{\frac{\theta}{2}p_{max}}=\\
&=&\cos(\theta) \gamma_j-(-1)^j\sin(\theta)\gamma_1...\hat\gamma_j...\gamma_{2n},\nonumber
\ee
if $n$ is odd, or
\be\label{def:actiontransf2}
e^{-\frac{\theta}{2}p_{max}} &\gamma_j& e^{\frac{\theta}{2} p_{max}}=\nonumber\\
&=&\cos(\theta) \gamma_j-i(-1)^j\sin(\theta)\gamma_1...\hat\gamma_j...\gamma_{2n},\nonumber
\ee
if $n$ is even. The hat over a Majoranas means that the Majorana has been removed.
The reader can note that, in the two cases, both $\gamma_1...\hat\gamma_j...\gamma_{2n}$ and 
$i\gamma_1...\hat\gamma_j...\gamma_{2n}$ behave as Majorana fermions, since they are hermitian and they square to $+1$.
It is quite evident that, via the definition (\ref{def:NLT}), this subgroup of transformations is exactly the $U(1)$ normal subgroup of LCG.
%, since it commutes with all the other transformations in NLT.
This is not surprising, since $p_{max}$ is the total local parity operator ($P_L$ in Appendix~\ref{app:Wenger}), as mentioned previously.
This normal subgroup is the generalized version of the {\it Hodge rotation} (\ref{eq:HodgeN2}) introduced in the case $n=2$.
This also explain our choice for its name: applying this transformation
to the basic elements of the algebra $\gamma_1,\gamma_2,...,\gamma_{2n}$ we obtain new Majoranas $\mu_1,...,\mu_{2n}$ as linear
combinations of the original generators of the Clifford algebra (row containing the single Majoranas) and their Hodge dual.

Our results make the use of canonical transformations conveniently easy, since (\ref{def:NLT})
is a quite neat formula and the form of all the generators has been found (\ref{defL})-(\ref{alfabeta}).
Moreover they also make clear the rationale behind the existence of the canonical transformations, showing how they can be understood in terms of Majoranas,
as a combination of the inequivalent sets of well defined Majorana fermions (or emergent\cite{Lee:2013aa,WilczekMajoranaErice} Majorana fermions) inside the full Clifford algebra.

In conclusion:
\begin{enumerate}
\item we proved that the set $\mathcal{T}^{(n)}$ obtained from the elements of the Clifford algebra generated by $2n$ Majoranas as in (\ref{defT}), closes to the Lie algebra $su(2^n)$;
\item we demonstrated that the set of antihermitian operators $\{p_i\}^{(n)}$, defined using the even elements of the Clifford algebra generated by $2n$
Majoranas, can be used via definition (\ref{def:NLT}) to generate the continuous part of the Local Canonical Group LCG;
\item we showed that the non-linear canonical transformations inside LCG, generated according to (\ref{def:NLT}), can still be interpreted as
mixing the rows of odd elements of the Clifford algebra, i.e., they can be thought as if they generate a new set of Majoranas starting from the inequivalent
sets that can be defined inside the Clifford algebra;
\item we identified a convenient form for the set of generators of the LCG group in an
Hilbert space of arbitrary dimension $2^n$. In analogy with the $\mathcal{H_H}$ case, the three subalgebras corresponding to the
three continuous subgroups of LCG are the two $\mathcal{L}^{(n)}_\alpha$ (ISH) and $\mathcal{L}^{(n)}_\beta$ (SSH), both isomorphic $su(2^{n-1})$ algebras, and
the element $p_{max}$, generating the $U(1)$ subgroup of the {\it Hodge rotation}, which is the generalization of the non-linear transformation found in the $\mathcal{H_H}$ case.
\end{enumerate}
The {\it crucial difference between the general case and the the case $n=2$ is that the Hodge rotation
is not anymore the only non-linear one}; it is instead a very peculiar non-linear transformation among many others,
which are contained within $\mathcal{L}^{(n)}_\alpha$ and $\mathcal{L}^{(n)}_\beta$.
Of course, to complete the set of all the possible canonical transformations in LCG, one should also consider
the discrete transformation introduced in Sec.~\ref{sec:sec1example},
which do not hide difficulties as we mentioned previously.
A summary of the different algabras and their relations with each other is give in Tab.~\ref{tabsummary}.

We would like to mention an important side product of all this formal construction, which may otherwise go unnoticed, pointing
the attention to the algebra $\mathcal{L}^{(n)}_{1/2}$. Such an algebra has the structure of an $su(2^{n-1})$ Lie algebra and it can be used to generate
the entire $SU(2^{n-1})$ group. What we want to stress is the fact that the $su(2^{n-1})$ elements in $\mathcal{L}^{(n)}_{1/2}$ are represented in terms
of Majoranas and this makes extremely easy to work with the elements of the algebra.
In physics literature many different representations of Lie algebras (in particular $su(n)$ algebras) have been elaborated.
The bosonic representation of $su(n)$ algebras (see Ref.~\onlinecite{Klein:1991aa} for a comprehensive review) is probably the most known, in particular thanks
to its vast use in the study of spin systems via the Holstein-Primakoff and Schwinger mappings\cite{Holstein:1940aa,SchwingerBosons,Auerbach}.
Also fermionic representations are possible (see Ref.~\onlinecite{Iachello} and references therein); for example the known Jordan-Wigner transformation\cite{Jordan:1928tg},
effective in the study of 1d spin chains, allows to represent $su(2)$ operators in terms of fermionic creation and annihilation operators.
Even if based on Majorana fermions operators and not on standard ones, our representation is obviously
closer to this latter class of representations, rather then the former. However, with respect to the known representations, the one that we have elaborated is
extremely natural in the context of non-linear canonical transformations (where we use it) and also its simplicity should be seen as a valuable strength.

To give an idea of the operators belonging to $\mathcal{L}^{(n)}_\alpha$ and $\mathcal{L}^{(n)}_\beta$, it is convenient to study some examples.

{\it - The 4-dimensional Hilbert space $\mathcal{H_H}$:}
\be
p_{max}=i\gamma_1\gamma_2\gamma_3\gamma_4,
\ee
and the operators in $\mathcal{L}^{(2)}$ all the bilinears
\be
p_1=\gamma_1\gamma_2, \quad p_2=...
\ee
Therefore one obtains:
\be
\alpha_1&=&p_1+ip_{max}p_1=\gamma_1\gamma_2+\gamma_3\gamma_4,\nonumber\\
\beta_1&=&p_1-ip_{max}p_1=\gamma_1\gamma_2-\gamma_3\gamma_4,\\
\alpha_2&=&...\nonumber
\ee
Hence, in the case of $\mathcal{H_H}$, the algebras $\mathcal{L}^{(2)}_\alpha$ and $\mathcal{L}^{(2)}_\beta$
are the two $su(2)$ algebras of isospin and spin (up to proper normalizations and multiplicative factors
necessary to have hermitian operators and appropriate normalizations). This notation will be generalized later on.
%\news{The even-odd characteristic of $\alpha$ and $\beta$ is then clear and it is a consequence
%of the fact that the two set of operators gives a non-zero result only if applied on odd/even parity states (mha... i have to think to this... irrelevant anyway).}

{\it - The 8-dimensional Hilbert space $\mathcal{H_K}$:}
\be
p_{max}=\gamma_1\gamma_2\gamma_3\gamma_4\gamma_5\gamma_6,
\ee
the operators in $\mathcal{L}^{(3)}$ are the bilinears and the quadrilinears
\be
\gamma_1\gamma_2,...,\, i\gamma_1\gamma_2\gamma_3\gamma_4,...
\ee
So the operators in $\mathcal{L}^{(3)}_\alpha$ and $\mathcal{L}^{(3)}_\beta$ look like
\be\label{eq:su4operators}
\begin{split}
\alpha_1=\gamma_1\gamma_2-i\gamma_3\gamma_4\gamma_5\gamma_6,\\
\beta_1=\gamma_1\gamma_2+i\gamma_3\gamma_4\gamma_5\gamma_6.
\end{split}
\ee
Please note that in terms of the original fermionic operators these are {\it fourth order} operators,
so they generate {\it non-linear} (canonical) transformations. Both the algebras are isomorphic to $su(4)$ and
clearly there are three simultaneously diagonalizable operators in both of them:
\be
\gamma_1\gamma_2&\pm& i\gamma_3\gamma_4\gamma_5\gamma_6,\nonumber\\
\gamma_3\gamma_4&\pm& i\gamma_1\gamma_2\gamma_5\gamma_6,\\
\gamma_5\gamma_6&\pm& i\gamma_1\gamma_2\gamma_3\gamma_4.\nonumber
\ee
Also in this case one should multiply these operators by the imaginary unit to ensure hermiticity.
The quantum numbers associated with these operators can be used, for example, to label all the states of the Hilbert space. 

{\it - The 16-dimensional Hilbert space $\mathcal{H_A}$:}
\be
p_{max}=i\gamma_1\gamma_2\gamma_3\gamma_4\gamma_5\gamma_6\gamma_7\gamma_8,
\ee
in $\mathcal{L}^{(4)}$ are present the bilinears, quadrilinears and hexalinears
\be
\gamma_1\gamma_2,...,\,i\gamma_1\gamma_2\gamma_3\gamma_4,..,\,\gamma_1\gamma_2\gamma_3\gamma_4\gamma_5\gamma_6,...
\ee
So the operators in both $\mathcal{L}^{(4)}_\alpha$ or $\mathcal{L}^{(4)}_\beta$ look like
\be
\begin{split}
&\gamma_1\gamma_2\pm\gamma_3\gamma_4\gamma_5\gamma_6\gamma_7\gamma_8,&\\
&i\gamma_1\gamma_2\gamma_3\gamma_4\pm i\gamma_5\gamma_6\gamma_7\gamma_8.&
\end{split}
\ee
In this case $\mathcal{L}^{(4)}_\alpha$ and $\mathcal{L}^{(4)}_\beta$ have an $su(8)$ structure, which means that they contain 63 elements
and 7 of them commute among each other.

%So far we have seen how NLT can be separated into three different subalgebras, which seem good candidates as the subalgebras of the subgroups
%$SU(2^{n-1})$ and $U(1)$ in LCG.
%What is left to do is to understand is if the structure of NLT is exactly the same of LCG, so if this separation procedure indeed decomposes NLT correctly.
%In fact it may be that the representation (\ref{def:NLT}) does indeed
%generate an isomorphism between NLT and LCG, and that the set of generators $\mathcal{L}_\alpha$, $\mathcal{L}_\beta$
%and $p_{max}$, indeed generate two isomorphic groups $SU(2^{n-1})$ and the $U(1)$ subgroup.

\section{Examples of Non-linear analysis of the sces}\label{sec:applications}
In the previous section we completely characterized the structure of the group of non-linear canonical transformations,
showing that a complete control of these transformations is achieved via the Majorana fermions representation.
It is appropriate to provide the reader with a clue about how this knowledge can be helpful in the analysis of physically
relevant problems. We strongly believe that the availability of controllable and systematic methods based on the group of
non-linear canonical transformations may benefit the study of the SCES.

In general the use of transformations of the quantum coordinates is a central technique, often applied in condensed matter contexts.
A look to any modern condensed matter textbook (as for example Ref.~\onlinecite{Fazekas,Auerbach,AltlandSimons}) should show a great variety of these techniques,
which embrace for example slave bosons approaches\cite{Coleman:1984vn}, bosonization\cite{Delft:1998pj},
low-energy projection methods\cite{Eder:1997kl,Eder:1998fu,Sinjukow:2002bd,Schrieffer:1966mw,Nozieres:1974fv,Sigrist:1992qa,Hellberg:1991mq,Feng:1994ys,MatsStellanHUBBARD},
and many others. In particular this latter class is interesting to us, since it makes often use of some kind of non-linear transformation.
However an important feature that makes the canonical non-linear transformations different from many of the these known techniques is the fact that the dimensions
of the Hilbert space (as of course the Fock structure) is always preserved. The non-linear canonical transformations may be considered a subgroup of
the generalized Bogoliubov-Valatin transformations, which have been mathematically considered in the cases of systems with one and two fermionic modes\cite{Scharnhorst:2011aa}.
In the past years canonical non-linear transformations have been used successfully in some situations\cite{Harris:1967qf,StellanMele,StellanKONDO},
but unfortunately no comprehensive analysis of their structure (that instead we provide in our work) has never been done before.
We hope that, thanks to our analysis, it will become possible to make this tool accessible to the general community and to
provide an understanding of the fundamental rationale that connects the different known non-linear methods.
Indeed, in the light of the framework that we introduced, it is possible to {\it systematically} apply methods based on non-linear canonical transformations
in the context of the SCES.

We provided in other occasions\cite{ChristmaMajorana,MBzNilssonPREP,Bazzanella:2014aa,JohanKondoORIGINAL} examples where
the use of this technique produced evidence of its effectiveness. We will review briefly here some of the core ideas we developed, in order
to help the reader to understand the crucial nature of the ideas and results developed in the present manuscript.
Moreover we will add a short discussion about correlated hopping models, since
the appearance of correlated hopping terms typically takes place when non-linear transformations are used. Our discussion does not intend to be complete,
but its aim is to highlight some results that we believe may be useful for future analysis and interesting for some readers. 

\subsection{Hubbard model}
As first example we cite the results we obtained in Ref.~\onlinecite{MBzNilssonPREP}.
In that situation we used a powerful general feature of the non-linear canonical transformations:
the fact that they allow to turn an interacting term into a quadratic one.
This is possible since via a canonical transformation one can represent an interacting Hamiltonian in
terms of new ``correlated'' fermions able to capture the physics in a more transparent way.
This kind of approach often becomes much easier in terms of Majorana fermions, since the algebra
becomes simpler.

Let us consider a simplified example and take the {\it local part} $H_{loc}$ of a lattice Hamiltonian $H=H_{K}+H_{loc}$ in a system with three fermion species:
\be\label{exampleHloc}
H_{loc}&=&2(X-U) c^\dd c + 2(Y-U) f^\dd f + 2Z g^\dd g\nonumber\\
&&\qquad +4 U c^\dd c f^\dd f,
\ee
which means that the quantities $2(X-U)$, $2(Y-U)$ and $2Z$, play the role of the chemical potentials.
In this system the interaction among the fermions
affects only the $c,f$ sector, leaving the $g$ sector completely free.
Rewriting the previous Hamiltonian in terms of Majoranas one obtains
\be\label{particularexample}
H_{loc}&=&X(-i\gamma_1\gamma_2)+Y(-i\nu_1\nu_2)+Z(-i\gamma_3\nu_3)\nonumber\\
&&\qquad-\frac{U}{2}(\gamma_1\gamma_2\nu_1\nu_2),
\ee
where we omitted some additive constants, irrelevant for the discussion and we defined $c^\dd=(\gamma_1+i\gamma_2)/2$,
$f^\dd=(\nu_1+i\nu_2)/2$ and $g^\dd=(\gamma_3+i\nu_3)/2$.
Now we can apply the transformation generated by $i\gamma_1\gamma_2\nu_1\nu_3$, obtaining:
\be
\nu_1\rightarrow \cos(\theta)\nu_1+\sin(\theta)i\gamma_1\gamma_2\nu_3, \nonumber\\
\nu_3\rightarrow \cos(\theta)\nu_3-\sin(\theta)i\gamma_1\gamma_2\nu_1, \\
i\gamma_1\gamma_2\nu_1\rightarrow \cos(\theta)i\gamma_1\gamma_2\nu_1+\sin(\theta)\nu_3. \nonumber
\ee
This means that under this non-linear canonical transformation, $H_{loc}$ becomes
\be\label{miraclerotation}
%H_{loc}&=&X(-i\gamma_1\gamma_2)+Y(-i)(\cos(\theta)\nu_1+\sin(\theta)i\gamma_1\gamma_2\nu_3)\nu_2\nonumber\\
%&&\quad+Z (-i)\gamma_3(\cos(\theta)\nu_3-\sin(\theta)i\gamma_1\gamma_2\nu_1)\nonumber\\
%&&\qquad+i\frac{U}{2}(\cos(\theta)i\gamma_1\gamma_2\nu_1+\sin(\theta)\nu_3)\nu_2. \nonumber\\
H_{loc}&=&X(-i\gamma_1\gamma_2)+Y\cos(\theta)(-i\nu_1\nu_2)\nonumber\\
&&\quad-\frac{U}{2}\sin(\theta)(i\nu_2\nu_3)+Z \cos(\theta) (-i\gamma_3\nu_3) \nonumber\\
&&\qquad-Z\sin(\theta)\gamma_1\gamma_2 \gamma_3\nu_1-\frac{U}{2}\cos(\theta)i\gamma_1\gamma_2\nu_1\nu_2\nonumber\\
&&\qquad\quad-Y\sin(\theta)\gamma_1\gamma_2\nu_2\nu_3.
\ee
In the previous form, if originally $Z=Y=0$ and if $\theta=\pi/2$ is chosen, then
\be
H_{loc}&=&-iX\gamma_1\gamma_2-i\frac{U}{2}\nu_2\nu_3.
\ee
Therefore the local interacting term is turned into a non-interacting one by the transformation and it can now be diagonalized,
making use of the fermionic operators $c,c^\dd$ and $\eta,\eta^\dd=(\nu_2+i\nu_3)/2$.
This means that, in some circumstances, in particular if the interaction involves only few of the fermion species, it is possible
to turn an interacting local Hamiltonian into a non-interacting one via a non-linear transformation.
Relaxing the hypothesis one can consider the more general case with $-U,\left|Y\right|,\left|Z\right|<< 1$. In this situation there is a controlled way to {\it trade a strong interaction term
with a number of weak ones}. In fact, since the sine and cosine take values only between $+1,-1$, it is clear from (\ref{miraclerotation}) that the
constant in front of the four fermion term $\gamma_1\gamma_2\nu_1\nu_2$ can be reduced arbitrarily, without causing an
equivalent increase of the coefficients in front of the other four fermion terms.

The example (\ref{particularexample}) is evidently quite particular, but it shows the potential benefits of non-linear canonical transformations.
The physical reason behind these benefits is that the correlation between the electrons can
cause the appearance of composite particles (such as singlets, for example) that are therefore more convenient degrees of freedom to describe the system with.
Since the correlations and the interactions (cor)relate the fundamental (original) fermions, it makes
sense to change the degrees of freedom that are used to describe the Hamiltonian. The non-linear canonical transformations permit
this goal to be achieved, keeping the fermionic language. In this sense it becomes possible to identify the ``best'' fermionic degrees of freedom
for the description of a system with specific characteristics, i.e., the fermionic degrees of freedom that are able to capture the
greatest part of the correlation. This also means that in a system described by the same Hamiltonian, but with different
values for the physical parameters (interaction, hopping, etc...) the ``best'' fermions may be different.

This approach works best in large local Hilbert spaces or when at least one of the fermionic species is not interacting with the others. Hence, to obtain results
in the Hubbard model one has to enlarge artificially the Hilbert space. We did this in Ref.~\onlinecite{MBzNilssonPREP} where we considered
on each site of the Hubbard model (in arbitrary dimensions and at half filling) also the quantum levels of one non-interacting auxiliary spinful fermion.
%This enlargement reminds in spirit the same operation performed in the DMFT, DMRG or holographic approaches.
In terms of Majoranas the Hamiltonian (in 1d for notational convenience) reads
\be\label{HubbardJohan}
H=-\frac{t}{4}\sum_{r,a} i\gamma_{r,a}\gamma_{r+1,a}-\frac{U}{4}\sum_{r}\gamma_{r,1}\gamma_{r,2}\gamma_{r,3}\gamma_{r,4},
\ee
where $\gamma_a$ with $a=1,..,4$ indicate the four different Majoranas constituting the original fermions, while the $\mu_a$ that form the
auxiliary fermions do not appear, because they do not interact with the original fermions and they do not hop from site to site. 
Since these fermions do not interact with the original ones, it is evident that their presence does not chance the quantum problem\footnote{This is evidently true
for the quantum average of any physical operator. Some caution must be payed in the definition of entropy.}.
Instead, what changes is the number and the kind of non-linear transformations available.
Thanks to the results of Sec.~\ref{sec:section1} we proved the existence of two (and only two) non-linear transformations $S_1$ and $S_2$ that were consistent with the $SO(4)$
symmetry of the system at half-filling\cite{YANG:1990aa} and time reversal symmetry.
This two parameter transformation, realized according to (\ref{def:NLT}) by the unitary operator $V=e^{i(\theta_1S_1+\theta_2 S_2)/2}$,
mixes the original eight Majoranas $\gamma_a,\mu_a$, generating a new set of Majoranas.
Written in terms of these new degrees of freedom, the Hubbard Hamiltonian (\ref{HubbardJohan}) changes its form.
For example the Hubbard interaction term $P_1=\gamma_{r,1}\gamma_{r,2}\gamma_{r,3}\gamma_{r,4}$ becomes:
\be\label{Hubbardinteraction}
VP_1 V^\dd=A_0(\theta_1,\theta_2) P_1+A_1(\theta_1,\theta_2) H_{12}+...,
\ee
where $H_{12}$ is a sum of bilinear Majorana operators, the letters $A_j(\theta_1,\theta_2)$ indicate simple trigonometric functions and the dots summarize
the presence of a few four- and six-order Majorana terms that are irrelevant for the present discussion.
We then took the Hubbard Hamiltonian in this new form and we studied it with a mean-field theory,
using a two spinful free-fermion Hamiltonian to generate the variational ground state, as function of three variational mean-field parameters
$t_1$, $t_2$ and $\lambda$:
\be
\begin{split}
H_{MF}=-\sum_{r,\sigma} &\left(t_1 a^\dd_{r,\sigma} a_{r+1,\sigma}+t_2 b^\dd_{r,\sigma} b_{r+1,\sigma}+h.c.\right)&\\
&+\lambda \sum_{r,\sigma} \left(a^\dd_{r,\sigma}b_{r,\sigma}+h.c.\right).&
\end{split}
\ee
Since part of the Hubbard interaction (\ref{Hubbardinteraction}) is quadratic in the Majoranas (thus in the $a_\sigma$ and $b_\sigma$ operators),
its mean-field analysis will contain an exact part of the (original) fermion correlation that cannot be captured by the standard mean-field approximations.

To find a candidate ground state we looked for the lowest energy local minima of the functional
$E(t_1,t_2,\lambda,\theta_1,\theta_2)=\langle 0_{MF}| H |0_{MF}\rangle$ into this 5d variational space,
where the dependence on $t_1,t_2,\lambda$ enters explicitly into the definition of the candidate ground state $|0_{MF}\rangle$, while the $\theta_1,\theta_2$ dependence
is explicit into the form of the Hamiltonian $H$. 
Thanks to this procedure it is clear that we greatly enlarge the variational space on which we can perform our mean-field study, since we can also optimize the angles of the non-linear transformation $V$,
and our result is still an upper bound on the ground state energy.
%In this way we can capture part of the correlation that is missed by the usual mean-field techniques, since the partial diagonalization shown in (\ref{Hubbardinteraction}) permits us to (partially)
%include the exact value of the correlation into our calculation.

Among many interesting results, we found at high U a ground-state solution that gives a {\it free-particle mean-field description of the paramagnetic Mott insulator},
with a variational energy that is at least as good as the one obtained using the Hubbard I approximation.
A more detailed and complete discussion can be found in the cited paper\cite{MBzNilssonPREP}, where we also point out that our scheme allows to study at mean-field
level also the metal-Mott paramagnetic insulator transition.

\subsection{Kondo lattice model}
In Ref.~\onlinecite{ChristmaMajorana,Bazzanella:2014aa,JohanKondoORIGINAL} we used our framework to study the Kondo lattice model (KLM).
We applied a non-linear transformation to create an {\it exact mapping} between the standard representation of the KLM,
in terms of fermion and spin operators, into a different one in terms of three spinless fermions.
What we discovered was that via a non-linear transformation it was possible to ``complete'' the Schrieffer-Wolff transformation\cite{Schrieffer:1966mw,Sinjukow:2002bd};
this transformation permits to map the infinite interaction limit of the Periodic Anderson Model (PAM), described in terms of
two spinful fermionic local degrees of freedom $c_{c,\sigma}$ and $f_\sigma$, into the KLM that can be seen as the low energy sector of the PAM and that is described in terms
of local impurity spins $\mathbf{S}_f$ and conduction electrons $c_\sigma$.
In terms of Majoranas the representation of the PAM Hamiltonian needs eight Majoranas per site (four for each spinful fermion),
while the KLM one needs seven of them: four for the spinful fermion ($\gamma_a$ with $a=1,...,4$) and three for the local 1/2-spin ($\mu_b$ with $b=1,2,3$).
We discovered that one of these seven Majoranas is {\it redundant and can be removed from the Hamiltonian via
a non-linear transformation}. This was realized exchanging one of the $\gamma_a$ Majoranas (for example $\gamma_4$) with the composite (emergent)
Majorana $i\mu_1\mu_2\mu_3$.
In this way the low energy sector of the PAM, i.e.~the KLM, can be described in terms of six Majoranas only ($\gamma_a$, $\mu_a$ with $a=1,2,3$) or equivalently
in terms of three fermions $c=(\gamma_1-i\gamma_2)/2$, $g=(\gamma_3-i\mu_3)/2$, $f=(\mu_1-\mu_2)/2$.
Our non-linear mapping can be summarized in terms of the original degrees of freedom as:
\be
c^\dd&=&c^\dd_{c,\ua} , \label{eq:c-cgfmap}\nonumber\\
g^\dd&=&-\frac{1}{2}\left[c^\dd_{c,\da}+c_{c,\da}+(c^\dd_{c,\da}-c_{c,\da})2S^z_f\right], \label{eq:g-cgfmap}\\
f^\dd&=&-i(c_{c,\da}-c^\dd_{c,\da})S_f^+ \label{eq:f-cgfmap}.\nonumber
\ee
In terms of these new degrees of freedom, the original antiferromagnetic 1d Kondo Hamiltonian is mapped into a new form.
In particular the local interaction term
\be
\begin{split}
H_{J}=J \, \mathbf{S}_f \cdot \mathbf{S}_c,
\end{split}
\ee
with $J>0$, in terms of Majoranas becomes
\be
H_{J}&=&\frac{J}{8} \Big\{\left(i\gamma_1\mu_1+i\gamma_2\mu_2+i\gamma_3\mu_3\right)+\\
&& +\Bigl(\gamma_2\mu_2\gamma_3\mu_3+\gamma_1\mu_1\gamma_3\mu_3+\gamma_1\mu_1\gamma_2\mu_2\Bigl)\Big\}\nonumber
\ee
This representation shows a natural symmetry of the local Kondo interaction between the three flavors of Majoranas, which is broken by the hopping terms.
In terms of the fermionic operators defined previously, the interaction becomes
\be
\begin{split}
H_{J}=&\frac{J}{4} \left(1-c^\dd c-f^\dd f-g^\dd g\right)+&\\
&+\frac{J}{2} \left\{ c^\dd c f^\dd f + i g^\dd g(c^\dd f-f^\dd c)\right\}.&
\end{split}
\ee
Clearly part of the interaction is quadratic in the new fermions and contributes negatively to the energy. Of course the quartic components,
together with other non-quadratic (correlated hopping) terms that appeared into the kinetic term, required the use of approximations.
Even if in this situation, as in the analysis of the Hubbard model, the non-linear transformation allows a partial diagonalization of the interaction (although in this case
we did not have the freedom to tune the magnitude of the non-linear transformation), it is not this feature the most interesting and effective one.
In fact, written in this different form, the Hamiltonian shows also a very non-trivial symmetry\cite{Bazzanella:2014aa}, which stabilizes the spin-selective Kondo insulator
phase\cite{Peters:2012rq,Peters:2012fc} in the phase diagram.
Indeed it can be shown that the quantity
\be
\begin{split}
\sum_r A_3(r)&=-\frac{1}{2}\sum_r \left(i\gamma_1\gamma_2+i\mu_1\mu_2\right)&\\
&=\sum_r \left(c^\dd c +f^\dd f -1\right),&
\end{split}
\ee
commutes with the Hamiltonian. Such a term, quite easily expressed in terms of the new fermionic degrees of freedom, assumes a very suspicious
form in the old coordinates. Indeed:
\be
A_3(r)=S^z_f(r)+c^\dd_{r,\ua} c_{r,\ua}-\frac{1}{2},
\ee
which was originally named as commensurability\cite{Peters:2012rq,Peters:2012fc} and identified as an important ingredient in the description of the system, using DMFT and DMRG techinques.
Thanks to our formalism, which makes this symmetry evident, a simple mean-field study of the three spinless fermion
Kondo lattice Hamiltonian permitted us to qualitatively and quantitatively characterize the spin-selective Kondo insulator phase,
while we captured some qualitative correct features in the rest of the phase diagram.

\subsection{Correlated hoppings}\label{corrhoppingmain}
In both the previous examples, the application of a non-linear transformation causes the appearance of correlated hopping terms.
This is a typical collateral effect of these transformations: indeed they can simplify the
local part of the lattice Hamiltonian $H_{loc}$ as shown in the example (\ref{exampleHloc}),
but they make the hopping term $H_K$ more involved, causing the appearance of {\it correlated hopping terms}.
The fact that non-linear canonical transformations can relate interacting systems to correlated hopping systems is indeed quite interesting,
since it is known\cite{Hirsch:2002aa,Hirsch:1989aa,Marsiglio:1990aa,Marsiglio:1994aa}
that correlated hopping terms can explain the appearance of unconventional superconductivity.
Although in our other works we dealt with these terms mostly using mean-field approximations, we believe it is convenient
in this circumstance to tackle them from a different perspective, in order to show the interplay that exist between correlated hopping
terms and non-linear transformations (and symmetries related to them). We do not aim to give a full analysis of this topic, but just
to point out the main features that we believe make this relation evident.

Let us use a simple example, considering the model Hamiltonian
\be\label{transHMAIN}
H= -t \sum_{r,\sigma} &&\left(c^\dd_{-\sigma} \tilde c_{-\sigma}+ \tilde c^\dd_{-\sigma} c_{-\sigma}\right)\cdot\\
&&\cdot\left\{1-2(c^\dd_{\sigma}c_{\sigma}+\tilde c^\dd_{\sigma}c_{\sigma})+4c^\dd_{\sigma}c_{\sigma}\tilde c^\dd_{\sigma}\tilde c_{\sigma}\right\},\nonumber
\ee
where we took the 1d model for convenience, but all of the following arguments work in an arbitrary number of dimensions on bipartite lattices,
and where we used the convention $c_\sigma=c_{r,\sigma}$
and $\tilde c_{\sigma}=c_{r+1,\sigma}$.
Clearly the local Hilbert space of such a model is $\mathcal{H_H}$, therefore the only non-linear transformation acting locally is:
\be
\begin{split}
c^\dd_\ua \rightarrow {c^\prime}^\dd_\ua=e^{2i\chi}c^\dd_\ua-\left(e^{2i\chi}-1\right)c^\dd_\ua c^\dd_\da c_\da, \\
c^\dd_\da \rightarrow {c^\prime}^\dd_\da=e^{2i\chi}c^\dd_\da-\left(e^{2i\chi}-1\right)c^\dd_\da c^\dd_\ua c_\ua.
\end{split}
\ee
In terms of Majoranas the non-linear transformation is generated by the operator $i \gamma_1\gamma_2\gamma_3\gamma_4$ and acts locally as
\be
\gamma^\prime_i=R_\chi^\dagger \gamma_i R_\chi\quad\text{with}\quad R_\chi=e^{i \frac{\chi}{2}\gamma_1\gamma_2\gamma_3\gamma_4},
\ee
which means
\be\label{candidateMajorana}
\gamma_{i} \rightarrow \gamma^\prime_{i} = \cos(\chi) \gamma_i + i \epsilon_{ijkl} \sin(\chi)  \gamma_j\gamma_k\gamma_l.
\ee
We will always consider transformations applied uniformly on the entire lattice, so
\be\label{definitionmappingMAIN}
R_{\chi}&=&\prod_r e^{i\chi\, (\gamma_{1,r}  \gamma_{2,r}\gamma_{3,r}\gamma_{4,r})},
\ee
where $\chi$ is independent on $r$, although also more general transformations may generate interesting results.
Choosing $\chi=\pi/2$ it is easy to see that this non-linear transformation turns (\ref{transHMAIN}) into
\be\label{freeMAIN}
H^\prime=-t \sum_{r,\sigma=\pm} \left({c^\prime}^\dd_{-\sigma} {\tilde c^\prime}_{-\sigma}+ \tilde c{^\prime}^\dd_{-\sigma} {c^\prime}_{-\sigma}\right).
\ee
This means that (\ref{transHMAIN}) is equivalent to the free Hamiltonian (\ref{freeMAIN}), if expressed in terms of the ``correlated'' fermions $c^\prime_\sigma$
defined by the non-linear transformation $\chi=\pi/2$.
Although the Hamiltonians $H$ and $H^\prime$ {\it describe the same physics} (so they must have the same free energy and
eigenstates, only written in terms of different fermionic degrees of freedom), they are completely different from an operative point of view. While $H^\prime$
can be diagonalized in terms of $c^\prime$ and ${c^\prime}^\dd$ operators, the Hamiltonian $H$ cannot and its study requires approximations, unless one
is clever enough to realize that it is just a free model in disguise.

In Appendix \ref{threefermions} we go further along this line, providing some more details concerning these models.
In particular we show that studying how the general Hamiltonian
\be\label{corrhoppMAIN}
H_{ch}(t_1,t_2,t_3)=\sum_{r,\sigma=\pm} &&\Bigl\{ \left(c^\dd_{-\sigma} \tilde c_{-\sigma}+ \tilde c^\dd_{-\sigma} c_{-\sigma}\right)\cdot\\
&& \cdot\left[t_1+t_2(n_\sigma+\tilde{n}_\sigma)+t_3n_\sigma\tilde{n}_\sigma\right] \Bigl\},\nonumber
\ee
transforms under the aforementioned non-linear transformation, it is possible to discover that
the 2d plane $t_1=-t_2$, in the three dimensional Hamiltonian space $(t_1,t_2,t_3)$, is left unchanged by the non-linear transformation.
This known\cite{Strack:1993fk,Ovchinnikov:1994aa,OVCHINNIKOV:1993uq,Arrachea:1994aa,Arrachea:1994ab,Airoldi:1995aa,Montorsi:1996aa,Aligia:2007aa,Montorsi:2008aa}
fact implies the existence of a quantity conserved by the non-linear symmetry, which in turns implies the conservation {\it of the parity of the number of the doublons},
which are hard-core bosons built as the bound state of two electrons of opposite spins,
and are a special case of the so-called $\eta$-paired states\cite{Yang:1989aa,Boer:1995ab}.
This result is consistent with the known results\cite{Arrachea:1994aa,Arrachea:1994ab}, which indeed identify also two special points
inside this 2d plane where the {\it number} of the doublons is conserved. We will return on this point in the next section, while the
interested reader can find a more details in the mentioned Appendix \ref{threefermions}, where we also
discuss briefly the possibility to generalize the analysis to a system with more fermion species.

\section{The Hyperspin and the Holon}\label{sec:noncan}

%The consequences of the results proved in Sec.~\ref{sec:section1} go beyond the mere identification of the canonical non-linear transformations.
%Indeed they open the door to different ways to look at quantum systems and at their challenges.
Another promising idea for the analysis of SCES, based on the results of Sec.~\ref{sec:section1}, is given by the concept of hyperspin.
Such an object has been used already in the analysis of the Hubbard model\cite{MatsStellanHUBBARD,KumarORIGINAL}, but thanks
to the identification of $\mathcal{L}^{(n)}_{1/2}$ it is possible to use it in more general cases. Although it did not play a crucial role in our previous works,
it allowed a better comprehension of the transformation used in Ref.~\onlinecite{Bazzanella:2014aa} and we believe
it is convenient to clearly develop its notion.

Given the Hamiltonian of a SCES, it is well known that the fermionic representation of the degrees of freedom
is not always the best one. In some cases it may be more meaningful to use, for example, spin degrees of freedom to describe the physics of a system,
as it happens in the $t-J$ or Kondo models. Typically it is possible to go from the fermionic description to a different one via a ``non-canonical'' transformation,
in the sense that the new degrees of freedom (spin-like) used to describe the system (or the low energy sector of a theory)
do not obey the fermionic anticommutation rules and are based on some fundamental symmetry of the Hamiltonian.
An example is the Schrieffer-Wolff (SW) transformation\cite{Schrieffer:1966mw,*Sinjukow:2002bd}, which connects the Anderson and Kondo models,
turning the original description in terms of conduction and impurity electrons into a description in terms of conduction electrons and impurity spins.
The convenience and the adequateness of this new representation is given by the fact that the Hilbert  subspace corresponding to the
impurity electron states is split in two parts highly separated in energy. The two low-energy degenerate (or quasi-degenerate) states
in this sub-space are easily described in terms of $SU(2)$ spins.
Clearly this description is convenient as long as the two states of the local impurity are degenerate, or almost degenerate if compared
with the other energy scales of the system. So in this case, as in general, it is the form of the Hamiltonian and the presence of
symmetries that makes one description preferable to another.

The use of symmetry (or algebraic) principles to choose the quantum coordinates for the representation of an Hamiltonian can significantly change our perspective on the problem.
Indeed this kind of approach is well known in physics, in particular in nuclear, particle and atomic context,
where concepts as spectrum generating algebras, dynamical symmetries and degeneracy algebras are largely used (see Ref.~\onlinecite{Iachello,BohmDS} for a comprehensive introduction).
These ideas had much less success in traditional SCES physics, where the infinite dimensionality of the full Hilbert space makes them less appealing.
Nevertheless an analysis of SCES Hamiltonians using symmetry and algebraic methods can still be a valuable option. The aforementioned Schrieffer-Wolff transformation
is an example of that, since the procedure identifies in the low energy limit a local $SU(2)$ degeneracy algebra that we interpret as a local spin degree of freedom.
Evidently, analysis of these kind assume the identification of (in general complicated) subalgebras on which the representation of the Hamiltonian can be based;
therefore they assume the definition of sets of spin-like operators that obeys the particular symmetries of the degenerate space. This identification must
be done starting, in the most standard cases, from a purely fermionic interacting Hamiltonian.
The determination of these spin-like operators and of their connection to the original fermionic operators is clearly not always easy.
Our characterization of the operators belonging to LCG and the use of the Majorana fermion representation helps in this sense, since many possible
spin-like algebras that can be used in the description of the system are contained into the LCG generators set.
In this sense we claim that the results of Sec.~\ref{sec:section1}, integrated by the arguments that we will highlight in this section, may help
in making these algebraic methods more easy to handle and interpret in the SCES context.

We mentioned some of these concepts in the Appendix A of our previous work\cite{Bazzanella:2014aa}. Here we will re-formulate some parts
of that discussion as examples, avoiding useless redundancies, and then we will provide a generalization of the main concepts.

\subsection{Example: the four dimensional Hilbert space}

As we have seen, the local Hilbert space $\mathcal{H_H}$ can be represented as a (not uniquely defined) Fock space.
However, all the basis states can also be represented as the tensor product of two states,
belonging respectively to the Fock space of the holon (represented by $h^\dd$, $h$) and
to the Hilbert space of a spin-like (hyperspin) degree of freedom (represented by $\mathbb{\vec S}$). The map is summarized in Table \ref{tab:Kumarmap}.
Immediately the reader will note that the hyperspin corresponds to the spin if projected on the subspace where one
holon present (i.e., on the subspace with an odd number of fermions) or charge isospin otherwise.
It must be stressed that in this context the term holon does not take the same meaning as in some other situations, as it happens for example in the cuprates literature\cite{Phillips:2010aa}.
In that context the holon indicates a vacancy in an electron system and consequently it bears information about the total local charge and it has bosonic character.
In our context instead it represents a fermionic particle that carries the information only about the parity of the local fermion number.

A deeper understanding of this non-canonical transformation of  the Hilbert space
(i.e., of the representation of the quantum degrees of freedom of the systems)
is obtained thanks to the Majorana representation, as suggested in Ref.~\onlinecite{KumarORIGINAL} and reviewed in Ref.~\onlinecite{Bazzanella:2014aa}.
The pivotal role is again played by the fact that the three composite object $\gamma_0=i\gamma_1\gamma_2\gamma_3$ is a well defined Majorana fermion.
Therefore it can be used together with the forth Majorana $\gamma_4$ to build a fermion operator
\be\label{defholon}
h^\dd=\frac{\gamma_0+i\gamma_4}{2}.
\ee
The reader should pay attention to the {\it very different nature} of this operation, with respect to those that led to (\ref{StellansNL}), for example.
This fermionic operator cannot be obtained via a rotation of the lines (\ref{Hubbard:first}) and (\ref{Hubbard:third});
moreover this procedure does not preserve the form of the Clifford algebra, since it splits it into two components:
a first (even dimensional) Clifford algebra, with two Majoranas $\gamma_4,\gamma_0$;
and a second (odd dimensional) one with three Majoranas $\gamma_1,\gamma_2,\gamma_3$.
The first Clifford algebra is used to generate the holon fermionic operators $h^\dd,h$, while the second creates the three {\it hyperspin operators}:
\be\label{quasispin}
\mathbb{S}_1=-\frac{i}{2}\gamma_2\gamma_3,\quad \mathbb{S}_2=-\frac{i}{2}\gamma_1\gamma_3,\quad \mathbb{S}_3=-\frac{i}{2}\gamma_1\gamma_2.
\ee
%\news{CHECK NORMALIZATION! THESE ARE THE OPERATORS FOR $\gamma^2=1/2$.}
Evidently these operators fulfill\footnote{Some attention should be taken if one wants to compare our operators with the standard ones. Indeed, comparing (\ref{quasispin})
with the Pauli matrices representation of $su(2)$, one can discover that in our convention $\mathbb{S}_1,\mathbb{S}_2,\mathbb{S}_3$ correspond to $\sigma_x,-\sigma_y,\sigma_z$.
Evidently the sign in front of $\mathbb{S}_2$ is irrelevant, but it could make the notation too clumsy, therefore we decided to stick to our convention, rather than the standard one.
This of course applies also to the general case of $su(n)$ algebra.}
the commutation relations of an $su(2)$ Lie algebra, typical of spin operators\cite{Shastry:1997xz}; as anticipated, their interpretation
as spin, rather than isospin degrees of freedom, depends upon the holonic part of the quantum state. This can be seen immediately writing the spin-isospin operators in
the original two fermions representation and then switching to the new one. For example\footnote{The reader should pay attention that in this circumstance
we did not use the unreasonable definition (\ref{eq:conventionHubbard}), but the more natural $c_j=(\gamma_{2j}-i\gamma_{2j+i})/2$, so the formulas of (\ref{sec:sec1example})
must be considered with an extra minus multiplying all the $\gamma_3$ Majoranas.}:
\be
S_3=-i\frac{\gamma_1\gamma_2-\gamma_3\gamma_4}{4},\quad I_3=-i\frac{\gamma_1\gamma_2+\gamma_3\gamma_4}{4},
\ee
which become
\be\label{projectionS}
\begin{split}
&S_3=-\frac{i}{2}\gamma_1\gamma_2 \frac{1-i\gamma_0\gamma_4}{2}=\mathbb{S}_3 h^\dd h&\\
&I_3=-\frac{i}{2}\gamma_1\gamma_2 \frac{1+i\gamma_0\gamma_4}{2}=\mathbb{S}_3 (1-h^\dd h).&
\end{split}
\ee
%\news{CHECK NORMALIZATION! THESE ARE THE OPERATORS FOR $\gamma^2=1/2$.}
%So it is clear that on the states with one holon (which correspond to the states with one fermion), the $\mathbb{\vec S}$
%reduces to the spin operator \news{(shouldn't it be that the spin is the hyperspin.. etc...? No... well..)},
%while on the states with no holon (which correspond to the states with no or two fermions) it reduces to the charge-isospin.
The representation of the Hamiltonian in these terms has been used successfully in the study of the Hubbard\cite{MatsStellanHUBBARD,KumarORIGINAL} and t-J\cite{Feng:1994ys} models.
Of course these kind of transformations can also be used vice-versa, as was done for example by the
authors in the study of the Kondo lattice model\cite{JohanKondoORIGINAL,Bazzanella:2014aa},
as mentioned previously.

In this example we focused to the algebras and operators of $\mathcal{H_H}$, but
the concept of hyperspin and holon are straightforwardly generalized also to Hilbert spaces of larger dimensions, following the same recipe.
%starting from the original ones written in terms of conduction electrons and local impurity spins. \news{(Horrid... fix it!).}
%All these concepts will now be generalized.

\begin{table}
\caption{\label{tab:Kumarmap}Mapping, as introduced in Ref. \onlinecite{KumarORIGINAL,MatsStellanHUBBARD},
between the two different representations of the Hilbert space associated with a local spinfull electron. On the left
the spinor representation, given by the operators  $c_\da$, $c_\ua$ and hermitian conjugates;
on the right the representation given in terms of holon and Pauli operators.}
\begin{ruledtabular}
\begin{tabular}{rcr}
\qquad$|0\rangle$&  $\longleftrightarrow$ & $|0_h\rangle\otimes|\Downarrow\rangle$\qquad \,\\
\qquad $|\ua\da\rangle$& $\longleftrightarrow$ & $|0_h\rangle\otimes|\Uparrow\rangle$\qquad \,\\
\\
\qquad $|\ua\rangle$&  $\longleftrightarrow$ & $|1_h\rangle\otimes|\Uparrow\rangle$\qquad \,\\
\qquad $|\da\rangle$& $\longleftrightarrow$ & $|1_h\rangle\otimes|\Downarrow\rangle$\qquad \,\\
\end{tabular}
\end{ruledtabular}
\end{table}

\subsection{General definition of hyperspin}\label{sec:defL12}

The reader has probably noticed some familiar details, which were anticipated in Sec.~\ref{sec:section2math}:
the operators (\ref{quasispin}) belong to $\mathcal{L}^{(2)}_{1/2}$ if the arbitrary excluded Majorana fermion is $\gamma_4$;
the projectors $h^\dd h$ and $1-h^\dd h$ are $(1- ip_{max})/2$ and $(1+ ip_{max})/2$; the operators of spin and isospin in (\ref{projectionS}),
which belong to $\mathcal{L}^{(2)}_{\beta}$ and $\mathcal{L}^{(2)}_{\alpha}$, are obtained multiplying $\mathcal{L}^{(2)}_{1/2}$
by $(1\mp ip_{max})/2$.
The generalization of the concepts of holon and hyperspin is therefore straightforward.

The holon, defined combining one Majorana with its Hodge dual consistently with (\ref{defholon}), clearly maintains the same structure and meaning
independently upon the total number of fermions.
For example choosing $\gamma_{2n}$:
\be
\begin{split}
& h^\dd=\frac{\gamma_0+i\gamma_{2n}}{2},&\\
& \gamma_0=p_{max}\gamma_{2n}=\mathcal{I}(2n)\gamma_1\gamma_2 ... \gamma_{2n-1}.&
\end{split}
\ee
The structure of the holon is independent of the dimension of the Hilbert space and it always distinguishes
between the states occupied by an odd and by an even number of fermions.
The hyperspin instead changes with the
dimension $2^n$ of $\mathcal{H}$, since its algebra is given by $\mathcal{L}^{(n)}_{1/2}\simeq su(2^{n-1})$, obtained from the Majorana fermions used to build $\gamma_0$.
Beside these differences, the hyperspin can always be thought of as the sum of two components: a spin-like one (SSH) and an isospin-like one (ISH),
which are given by its projection on the subspaces with an odd (using $1-ip_{max}$) and even (using $1+ip_{max}$) number of fermions respectively.
We remind the reader that all this discussion is based on the local Hilbert space $\mathcal{H}$, i.e., the Hilbert space associate to a single site
of our system, where we set up the degrees of freedom that can be used to the study of the Hamiltonian of the infinite system.

These concepts are not merely mathematical, but they hide important and basic physical meanings. Typically
the quantum systems are described in terms of spin and orbital degrees of freedom; for example a sixteen dimensional
Hilbert space $\mathcal{H_A}$ is often described as the Fock space generated by four fermion species $c_\ua$, $c_\da$, $f_\ua$, $f_\da$.
This way to represent $\mathcal{H_A}$ is based on the identification of an orbital quantum number, given by the indices $c,f$, and a spin
quantum number, given by the indices $\ua,\da$, that can be used to label the quantum states of a fermion. This means that
the quantum states are labeled in terms of the algebra $SU(2)_{spin}\otimes SU(2)_{orbital}$, together with other quantum numbers as
the fermionic number (total charge). But this is a mere conventional choice. Indeed the classification of the states could be done, for example, in terms
of the $SU(8)$ algebra that embeds the spin-orbital semi-simple subalgebra. This means, focusing on the projection SSH and ISH of $\mathbb S$,
for which the eight dimensional Hilbert subspace form two IRREPs.
The basis states of such an IRREPs can be determined and labeled using the seven Cartan elements of $SU(8)$, but to do it and to put in correlation
the fermion based spin-orbital representation with this $SU(8)$ hyperspin representation (information needed for any practical purpose),
one must know have the multi-fermion representation of the generators of $SU(8)$.
In Sec.~\ref{sec:section1} we provided this information, showing that a possible choice for such Cartan elements (within $\mathcal{L}^{(4)}_{1/2}$) is:
\be
&&\gamma_1\gamma_2,\,\gamma_3\gamma_4,\,\mu_1\mu_2,\nonumber\\
&&i\gamma_1\gamma_2\gamma_3\gamma_4,\, i\gamma_1\gamma_2\mu_1\mu_2,\, i\gamma_3\gamma_4\mu_1\mu_2,\nonumber\\
&&\gamma_1\gamma_2\gamma_3\gamma_4\mu_1\mu_2,\nonumber
\ee
where we considered 8 Majoranas $\gamma_1$, $\gamma_2$, $\gamma_3$, $\gamma_4$, $\mu_1$, $\mu_2$, $\mu_3$, $\mu_4$
and arbitrarily removed $\mu_4$ to generate $\mathcal{L}^{(4)}_{1/2}$ from $\mathcal{L}^{(4)}$;
the previous operators are Cartan elements of $\mathbb S$
and they must be multiplied by $(1- ip_{max})/2$ to give the ones of SSH (i.e., $\mathcal{L}^{(4)}_\beta$).
Of course, since we are interested in describing observables, one should also multiply the operators by $i$, in order to obtain hermitian operators
both in the case of $\mathbb S $ and of SSH (or ISH).

The crucial point is that {\it as long as no Hamiltonian is defined} the difference between the two representations based on $SU(2)_{spin}\otimes SU(2)_{orbital}$ and $SU(8)$
is purely academic,
but when a specific Hamiltonian is defined, then we should expect that the physics,
via the subsequent possible breaking of the most general local (dynamical) symmetries, indicates
uniquely the correct picture and therefore the most natural representation.
The Hamiltonian determines, only on the basis of its own local symmetries (interactions), what are the most natural subalgebras,
and therefore good local degrees of freedom, among the infinitely many possibilities.
Of course this does not exclude the possibility to impose the conservation of other symmetries
(such as time-reversal, charge conservation, etc...) excluding elements form the algebra of $\mathbb{S}$.

This way of tackling the problem (which is clearly in the spirit of the spectrum generating algebra techniques)
changes significantly the interpretation and the insight that one can have the quantum system.
Moreover also from a operational point of view the scenario changes. In fact, classifying the basis states in terms of smaller and smaller
subalgebras and choosing a representation of the quantum system (and of its Hamitlonian) based on these smaller subalgebras
makes more and more difficult the identification and understanding of effects that involve their correlation, which is based on higher symmetry groups.
A discussion about these problems can be found for example in Ref.~\onlinecite{Li:1998aa} and references therein.
These kind of situations, where the physics of the systems obeys symmetry groups that are higher than the ones implicitly defined
by the standard formalism, are becoming more and more common and experimental set ups have already been realized\cite{Keller:2014aa}.

As we said at the beginning of this section, the study of quantum systems using more general symmetry groups is not a novelty in physics
and in particular the concepts that we discussed in the previous paragraphs can all be related to the theories of spectrum generating algebras,
dynamical symmetries and related topics\cite{Iachello}. Our claim is that, in the context of SCES, the use of such ideas is made much more easy
and natural in the Majorana fermion representation, since the knowledge of $\mathcal{L}^{(n)}$ allows one to think efficiently in these general terms,
starting from the most general symmetry group able to describe the Hilbert space and letting the Hamiltonian
determine what are the most appropriate subalgebras (and fermion degrees of freedom) to describe the system.
In the Majorana representation, in the light of the results of
Sec.~\ref{sec:section1}, it is possible to identify, define and easily handle these larger symmetry groups and algebraic structures. This helps the study
on both a practical level, simplifying enormously the non-commutative algebras of the huge Lie groups, and on a conceptual level, 
removing the (often unjustified) asymmetries in the treatment of different quantum numbers that the standard formalism demands.
It is the focus placed on the symmetries of the system and of the Hamiltonian that makes the holon-hyperspin representation
particularly interesting and hopefully more efficient in the study of some relevant systems.
%If the analysis of a system focuses on its symmetries, then the hyperspin representation
%permits easy identification of them and understanding of how the different interaction constants affects such symmetries.

%\onlinecite{Zhang:1997aa,Keller:2014aa,Li:1998aa,Kikoin:2012aa,YANG:1990aa,*Carmelo:2010aa,*Baeriswyl:1995aa,*Masumizu:2005aa,*Podolsky:2009aa}.

\subsection{Correlated hopping models in the hyperspin formalism}\label{sec.corrhopphyperspin}

To illustrate the convenience of the previous concepts, we propose here a discussion of the correlated hopping model analyzed previously
in terms of holon-hyperspin.

Let us consider a Hamiltonian of the type (\ref{corrhoppMAIN}) that is sent into itself by the non-linear transformation (\ref{definitionmappingMAIN}) with $\chi=\pi/2$.
One can rewrite the Hamiltonian in terms of holon and hyperspin operators via the following identifications:
\be\label{choichen2}
\begin{split}
&h^\dd=\frac{i\gamma_1\gamma_2\gamma_3+i\gamma_4}{2},\quad h=\frac{i\gamma_1\gamma_2\gamma_3-i\gamma_4}{2},&\\
&\mathbb{S}_1=i\gamma_2\gamma_3,\quad \mathbb{S}_2=i\gamma_1\gamma_3,\quad \mathbb{S}_3=i\gamma_1\gamma_2,&
\end{split}
\ee
where we assumed $c^\dd_\ua=(\gamma_1+i\gamma_2)/2$ and $c^\dd_\da=(\gamma_3+i\gamma_4)/2$ and where we changed
the normalization of the hyperspin operator for notational convenience.
Some straightforward algebra leads to the following representation of the Hamiltonian: %\news{(make t's consistent, credo siano $t_1$ e $t_3$ o $-t_1$)}
\begin{widetext}
\be\label{Hn2hyper}
H&=&\sum_r \Bigl\{\left(\frac{t_1}{2}+\frac{t_3}{4}\right)\left[\left(\mathbb{S}_1\tilde{\mathbb{S}}_1+\mathbb{S}_2 \tilde{\mathbb{S}}_2+\mathbb{S}_3 \tilde{\mathbb{S}}_3+1\right)
\left(h^\dd\tilde h+\tilde h^\dd h\right)\right]\\
&&\qquad\qquad\qquad+\frac{t_3}{4}\left[i\left(\mathbb{S}_1\tilde{\mathbb{S}}_2-\mathbb{S}_2\tilde{\mathbb{S}}_1\right) \left(h^\dd\tilde h-\tilde h^\dd h\right)
-\left(\mathbb{S}_3+\tilde{\mathbb{S}}_3\right)\left(h^\dd\tilde h+\tilde h^\dd h\right)\right]\Bigl\}.\nonumber
\ee
\end{widetext}
If $t_3=0$, the Hamiltonian has a global $SU(2)$ symmetry already mentioned and used in the literature\cite{Arrachea:1994aa}. 
The hyperspin representation makes it manifest and provides a neat way to understand it and make use of it.
%which is manifestly conserved since the hyperspin is based on the $SU(2)$ group.
%The hyperspins describe in a neat way the local symmetries mentioned in Ref.~\onlinecite{Arrachea:1994aa} and allow to use them in a very neat way,
%making very natural the procedure outlined in the cited literature.

The $t_3=0$ case is not the only $SU(2)$ symmetric one\cite{Arrachea:1994aa}.
In fact one can note that the choice (\ref{choichen2}) is not unique, but that an equivalent one can be obtained exchanging $\gamma_4$ with one of the
other three Majoranas. For example exchanging it with $\gamma_3$, which means using the transformation $\exp(\pi \gamma_3\gamma_4/2)$, one obtains the new definitions
\be\label{secondchoice}
&&h^\dd=\frac{i\gamma_1\gamma_2\gamma_4-i\gamma_3}{2},\quad h=\frac{i\gamma_1\gamma_2\gamma_4+i\gamma_3}{2},\nonumber\\
&&\mathbb{S}_1=i\gamma_2\gamma_4,\quad \mathbb{S}_2=i\gamma_1\gamma_4,\quad \mathbb{S}_3=i\gamma_1\gamma_2.
\ee
It is convenient to use also the following transformations together with the previous one:
$\exp(\pi\gamma_1\gamma_2/2)$ and  $\exp(i\pi\gamma_1\gamma_2\gamma_3\gamma_4/2)$.
The first is a normal rotation around the third axis of the hyperspin, while the second performs the transformation
$$
{h}^\dd\rightarrow i{h}^\dd \quad {h}\rightarrow -i{h}.
$$
All together they perform the following transformation:
\be
&{\mathbb{S}}_3  h\rightarrow h,\quad {\mathbb{S}}_3  {h}^\dd\rightarrow h^\dd&\nonumber\\
&\mathbb{S}_1\rightarrow \mathbb{S}_1(2{h}^\dd h-1),\quad \mathbb{S}_2\rightarrow \mathbb{S}_2(2{h}^\dd h-1),
\quad \mathbb{S}_3\rightarrow \mathbb{S}_3.\nonumber&
\ee
Applying this non-linear transformation on every other site, for example only on the tilde operators in (\ref{Hn2hyper}),
one discovers that the Hamiltonian with $t_3=-2t_1$ is mapped exactly into the Hamiltonian with $t_3=0$ and it is therefore $SU(2)$ symmetric.

One may wonder if similar situations can be found in multiband models. Such a question can be answered thanks to the results of Sec.~\ref{sec:section2math}
and in particular to the identification of $\mathcal{L}_{1/2}$.
Let assume that we are interested in finding an Hamiltonian that is $SU(4)$ symmetric in the case $n=3$,
analogous to the Hamiltonian (\ref{secondchoice}) with $t_3=0$ in the $n=2$ case.
We define our three fermions as in (\ref{threefermionsdef}).
We know that an $SU(4)$ hyperspin is defined in this case and it
is built on a three fermion species model. This means that the orientations of the $SU(4)$ hyperspin
distinguish between the four odd particle numbers states $|c\rangle,|f\rangle,|g\rangle,|cgf\rangle$, if one holon present,
and $|cf\rangle,|fg\rangle,|cg\rangle,|0\rangle$ if no holon is present.
The generators can be easily found identifying a correct $\mathcal{L}^{(3)}_{1/2}$ algebra. Deciding to use
the Majorana $\rho_2$ (\ref{threefermionsdef}) to form the holon, the $\mathcal{L}^{(3)}_{1/2}$ hyperspin algebra is given by the operators:
\be\label{su4hyperspin}
&&E_1=i\gamma_1\gamma_2,\quad E_2=i\gamma_1\mu_1,\quad E_3=i\gamma_1\mu_2,\quad E_4=i\gamma_1\rho_1,\nonumber\\
&&E_5=i\gamma_2\mu_1,\quad E_6=i\gamma_2\mu_2,\quad E_7=i\gamma_2\rho_1,\quad E_8=i\mu_1\mu_2,\nonumber\\
&&E_9=i\mu_1\rho_1,\quad E_{10}=i\mu_2\rho_1,\quad E_{11}=\gamma_2\mu_1\mu_2\rho_1,\nonumber\\
&&E_{12}=-\gamma_1\mu_1\mu_2\rho_1,\quad E_{13}=\gamma_1\gamma_2\mu_2\rho_1,\nonumber\\
&&E_{14}=-\gamma_1\gamma_2\mu_1\rho_1,\quad E_{15}=\gamma_1\gamma_2\mu_1\mu_2.
\ee
%We mention the fact that to substitute the operators that contain $\rho_2$ in a consistent way one can multiply them by $ip_{max}$;
The sign in front of $E_{12}$ and $E_{14}$ is chosen for future convenience; all the signs can be generated properly,
building a possible set for $\mathcal{L}^{(3)}_{1/2}$ and multiplying all the elements that contain $\rho_2$ by $ip_{max}$.
The holon operators are instead
\be\begin{split}
h^\dd=\frac{\gamma_1\gamma_2\mu_1\mu_2\rho_1+i\rho_2}{2},\\ h=\frac{\gamma_1\gamma_2\mu_1\mu_2\rho_1-i\rho_2}{2}.
\end{split}\ee
The operators (\ref{su4hyperspin}) can be used to build an $SU(4)$ symmetric Hamiltonian of the form:
\be\label{su4hamiltonian}
H_{su(4)}=\frac{t}{4}\sum_{r} \sum_{j=1}^{15}\left\{\left(E_j\tilde E_j+1\right)(h^\dd \tilde h+\tilde h^\dd h)\right\}.
\ee
This Hamiltonian is manifestly $SU(4)$ symmetric, since the term $\sum_{j=1}^{15} E_j\tilde E_j$, commutes with all the generations of $SU(4)$:
\be
\left[E_i+\tilde E_i , \sum_{j=1}^{15} E_j\tilde E_j\right]=0,\forall i.
\ee
The constant factor in $E_j\tilde E_j+1$ has been fixed consistently with (\ref{Hn2hyper}),
but it is not a strict requirement. A Hamiltonian that has exactly the same $SU(4)$ symmetric form has been studied for example in
Ref.~\onlinecite{Li:1998aa}, as can be seen identifying the two $su(2)$ commuting subalgebras $\{s_i\}$ and $\{t_i\}$ with
$\{E_1,E_4,E_7\}$ and $\{E_8,E_{13},E_{14}\}$ respectively. However in that situation the Hamiltonian was built in a $n=4$ model and the orientation
of the hyperspin was used to distinguish between the four different single particle states.

Given this result the other $SU(4)$ symmetric point can be found as in the $n=2$ case, inverting $\rho_2$ with one of the other Majoranas.
Moreover this Hamiltonian can be written in fermionic form, expanding the shorthand notation (\ref{su4hamiltonian}) in terms of Majoranas
and rewriting it using the original $c$, $f$, $g$ fermions, as done in Appendix \ref{threefermions}.

\section{Conclusions}
We have analyzed the structure of the group of canonical transformations of a lattice system with $n$ fermion species.
We have shown that using the Majorana fermion representation it becomes simple to determine the generators of the continuos part of the
canonical transformation group. In particular we proved how the elements of the Clifford algebra generated by $2n$ Majoranas
can close to Lie algebra, if properly multiplied by imaginary units. 
We have been able to characterize the different Lie subalgebras that compose the canonical group, providing also in this case a simple form for the generators
and showing how the Majorana formalism is extremely convenient if one wants to work with non-linear transformations.
We have also shown why the use of the canonical non-linear transformations can help in the analysis of the SCES, allowing for the
definition of degrees of freedom that contain more of the correlated physics of the system, or helping in the determination of
symmetries that may otherwise be difficult to discover. In this context the concept of holon and hyperspin has been defined, with respect to the algebraic structures
previously introduced.

As example of the usefulness of this approach in the context of the SCES we reviewed briefly some applications that have been
explored in previous works. Moreover we provided a brief discussion about correlated hopping models, since these kind of terms
appear naturally using our framework. In this context the identification of the algebras formed by the generators of the non-linear transformations provides
a powerful tool for the development of artificial Hamiltonians with specific symmetry properties and for the application of concepts such as dynamical symmetries, spectrum generating
and degeneracy algebras to the analysis of SCES.

In this work we focused our attention only on the formalism itself. We hope that this manuscript may become a valuable guide to anyone interested
to use non-linear methods in the context of the SCES.

\appendix

\section{The canonical group LCG}\label{app:Wenger}
In this appendix we will review the arguments originally presented in Ref.~\onlinecite{StellanMele}, without focusing on the small four-dimensional local Hilbert space.
Although many of the arguments were already outlined in the cited literature, we are going to generalize them to an arbitrary number of fermion species
and to organize them in a form that is convenient for the development of the main part of the paper.

Given a generic Hilbert space $\mathcal{H}$ of dimension $2^n$, it is clear that the most general non-trivial unitary transformation acting on it belongs to the group $SU(2^n)$,
since $\mathcal{H}$ is a complex vector space of dimension $2^n$ and the unitary transformation acts on it.
Given a generic element $U$ of $SU(2^{n})$, represented by the standard irreducible $2^n\times 2^n$ matrix $x$,
the procedure outlined in Ref.~\onlinecite{StellanMele} permits to represent it as a polynomial $P(x)$ of normal ordered creation and annihilation
operators, making use of the $2^n \times 2^n$ Wenger's matrix $m$:
\be\label{Wengertrans}
P(x)=Tr(xm).
\ee
It follows that $P(x)P(y)=P(xy)$ and $P(x)^\dd=P(x^\dd)$, which implies that the representation $P(x)$ of $SU(2^n)$ is faithful.
This can be proved by reduction ad absurdum. Let us assume that there exists a continuous set $\{x_a\} \in SU(2^n)$
such that $P(x_A)=1, \forall x_A\in\{x_a\}$. Thus, taken a generic element $x_B\in SU(2^n)$ and $x_A\in\{x_a\}$,
one should have $P(x_B x_A x_B^\dd)=P(x_B)P(x_A)P(x_B)^\dd=1$, therefore {\it for all} $x_B\in SU(2^n)$ one must have $x_B x_A x_B^\dd=x_c\in \{x_a\}$.
This means that $\{x_a\}$ is an invariant subgroup of $SU(2^n)$. Since $SU(2^n)$ is a simple Lie group, it contains no invariant subgroup, but the trivial one;
therefore $\{x_a\}$ contains only the identity. It is clear that this demonstration can be easily adapted to a discrete set $\{x_a\} \in SU(2^n)$ also.
The implication $P(x)=1\Rightarrow x=\mathbb{1}$ means that the representation is faithful. In fact, assuming that $P(x_C)=P(x_B)$
with $x_C\neq x_B$, then $1=P(x_B)P(x_C^\dd)=P(x_Bx_C^\dd)$, which means $x_C=x_B$, which is against the initial assumptions.
The fermionic representation of the generic element $U$
is more handy than the usual matrix representation $x$, when one is dealing with operators, since it permits to use the second quantization formalism in its full glory.

Automatically the previous properties permits one to prove that, independently upon the value of $n$,
the commutation relations $\{c^\dd_i,c_j\}=\delta_{i,j}$, with $i,j \in \{1,...,n\}$ are preserved by any unitary transformation $U$ that belongs to $SU(2^n)$.
In fact, taking a $2^n \times 2^n$ matrix representation $x\in SU(2^n)$, remembering that $x x^\dd=\mathbb{1}$,
and that its action on an operator is
\be\label{eq:actionU}
\begin{split}
&c_i \rightarrow c^\prime_i={P(x)}^\dd c_i P(x), &\\
&{c}^\dd_i \rightarrow {c^\prime}_i^\dagger ={P(x)}^\dd {c}^\dd_i P(x),&
\end{split}
\ee
it is then clear that the group of transformations $P(x)$ is composed (with the exception of $x=\mathbb{1}$) of only non-trivial canonical transformations,
which means that $\{c^\prime_i,{c^\prime}^\dd_j\}=\delta_{ij}$ and $c_i^\prime\neq c_i$ for at least one $i$.
To reach this conclusion one can note that no polynomial in $c_i,c_i^\dd$ can commute simultaneously
with all the creation/annihilation operators $c_1,...,c_n,c^\dd_1,...,c^\dd_n$, with the exception of the trivial scalar, which is in fact generated by $x=\mathbb{1}$, given
the faithfulness of the representation $P(x)$.
En passant we remind the reader that, since the transformation $U$ is acting on the operators as (\ref{eq:actionU}) and since a unitary transformation cannot change any matrix element,
then $U$ is also acting on the basis states $|\Psi_i\rangle$ of $\mathcal{H}$,
%then the action of $SU(2^n)$ on the basis states $|\Psi_i\rangle$ of $\mathcal{H}$ is also given by $U$,
although a better representation of the group of
transformations is given by the matrix form $x$, since we typically represent states of $\mathcal{H}$ as elements in a complex vector space.
%and not as polynomials of creation/annihilation operators acting on a vacuum state.

As we mentioned in Sec.~\ref{sec:section1}, we require the fulfillment of the generalized constraint (\ref{eq:StellanConstraint}):
\be
\{c^\dd_i(r),c_j(r^\prime)\}=\delta_{ij}\delta_{rr^{\prime}},
\ee
to define the local group of canonical transformations LCG $\subseteq SU(2^n)$, which acts on the local Hilbert spaces.
Always following the arguments of Ref.~\onlinecite{StellanMele}, it turns out that LCG is composed by all the
$P(x)$ built up using only linear combinations of an even number of fermionic operators $P_{even}(x)$.
Given this information, the structure of LCG can be understood immediately.
In fact it is clear that $P_{even}(x)$ contains {\it all} the operators that commute with the local parity operator $P_L$.
Reasoning backwards, it also means that we can define a subgroup of LCG as the subgroup of transformations $U$ in $SU(2^n)$ that commute with $P_L$.
For reasons that will become clear later, we call this subgroup the {\it continuous part of LCG}.
%This means on the one hand that, \todilo{if $U$ is represented as $P(x)$, then the polynomials are formed} only by the sum of even fermionic operators;
Immediately one can understand that if a $U$ which belongs to this continuous part of LCG is represented as a matrix $x$
acting on the basis states of $\mathcal{H}$, then such a matrix must be block diagonal, with
two blocks of equal size representing the action of $U$ on the even parity states and on the odd parity states.

These even/odd subspaces are Hilbert spaces of dimension $2^{n-1}$. It is therefore evident that it is possible to act with
two separate groups $SU(2^{n-1})$ on these separate subspaces. Since the constraint (\ref{eq:StellanConstraint}) is fulfilled
by the demand that $U$ does not mixes the two sectors, then all the transformations that act separately on the two sectors
belong to the continuous part of LCG. So we must have that at least
\be\label{susu}
SU(2^{n-1})\otimes SU(2^{n-1})\subseteq \text{ LCG } \subseteq SU(2^n).
\ee
It is convenient to point out here that the elements of LCG, which can be written down as $P_{even}(x)$ and form a continuous subgroup,
admit also a more compact representation. In fact the elements $x\in SU(2^{n-1})$ can be represented as $\exp (-t_i \mathbf{g}_i)$,
with $t_i\in \mathbb{R}$ and $\mathbf{g}_i$ antihermitian generators of the Lie algebra $su(2^{n-1})$.
Considering that via (\ref{Wengertrans})
the $x$ can be represented as $P_{even}(x)$, it is evident that the generators $\mathbf{g}_i$ must also be expressible as antihermitian
combinations of objects built using {\it only an even} number of fermionic operators.
Thinking in these terms, one realizes that a one dimensional subgroup is still missing in (\ref{susu}). In fact by construction,
among the transformations that belong to $SU(2^{n-1})\otimes SU(2^{n-1})$,
there is no non-trivial transformation that can commute with all the the elements of $SU(2^{n-1})\otimes SU(2^{n-1})$. But,
inside LCG, there exists a continuous group of transformations with this property, built up using the parity operator
$P_L$ itself to create an antihermitian generator $\mathbf{g} (P_L)$.
The group of transformations obtained in this way evidently belongs to LCG, but not to the $SU(2^{n-1})\otimes SU(2^{n-1})$ subgroup.
Moreover it can be defined to be isomorphic to $U(1)$ by construction.
%Therefore it must exist another continuous, at least one dimensional, subgroup of transformation in $SU(2^n)$
%that commutes with all the members of $SU(2^{n-1})\otimes SU(2^{n-1})$.
So the structure of the continuous part of LCG that we identified so far is
%This group can only be $U(1)$ since
\be
SU(2^{n-1})\otimes SU(2^{n-1}) \otimes U(1).
\ee
This is a maximal Lie subgroup\cite{Feger:2012fk} of $SU(2^n)$. Therefore, up to a finite number of discrete elements,
\be
\text{LCG} \simeq SU(2^{n-1})\otimes SU(2^{n-1}) \otimes U(1).
\ee
A careful analysis shows that there exist only one possible discrete transformations in LCG. This transformation exchanges particles-holes
of one fermion specie and has therefore the structure of a $\mathbb{Z}_2$ transformation, which inverts the parity of the system.
All the other transformation can be obtained combining this exchange with a continuous transformations, as mentioned in Sec.~\ref{sec:sec1example}.
We will not analyze this discrete transformation in detail, but simply justify later in the text the presence of only one $\mathbb{Z}_2$ transformation group.

\section{Lie algebras and Majorana operators}\label{qpp:Johansapproach}
In this section we are going to show discuss the relation between the Clifford algebra generated
by a set of Majoranas and the $su$ algebras.

Consider a set $\Gamma_{2m}$ of $2m$ Majoranas $\gamma_1,...,\gamma_{2m}$. It is well known\cite{Okubo:1991aa,WeyClassical}
that these operators admit representation as $2^m\times2^m$ matrices that can be obtained as the tensor product of $m$ matrices of dimension $2\times2$.
We chose the following representation:
\be
&&\gamma_1=\sigma_1\otimes\mathbb{1}\otimes ...\otimes \mathbb{1},\nonumber\\
&&\gamma_2=\sigma_3\otimes\mathbb{1}\otimes ...\otimes \mathbb{1},\nonumber\\
&&\gamma_3=\sigma_2\otimes\sigma_1\otimes \mathbb{1}\otimes  ...\otimes \mathbb{1},\nonumber\\
&&\gamma_4=\sigma_2\otimes\sigma_3\otimes \mathbb{1}\otimes  ...\otimes \mathbb{1},\nonumber\\
&&...\nonumber\\
&&\gamma_{2m}=\sigma_2\otimes\sigma_2\otimes  ...\otimes \sigma_2\otimes\sigma_3,
\ee
which clearly yields to proper Majorana fermions.
Consider now the elements of the full Clifford algebra generated by these Majoranas, both those obtained multiplying together
and even number of Majoranas $\{\Omega_i\}^{(m)}$, but excluding the identity element, and those given by the multiplication of an odd
number of Majoranas $\{\Delta_i\}^{(m)}$, where we used the notation (\ref{originalelements}).
It is straightforward to check that there are in total $2^{2m}-1$ elements, all linearly independent.
Moreover if they are multiplied by proper imaginary units they are also all hermitean and define the sets $\{p_i\}^{(m)}$ and $\{d_i\}^{(m)}$ as in (\ref{randd}).
Therefore, in the matrix representation defined above, they provide a orthogonal basis for all the traceless Hermitean $2^m\times2^m$ matrices.
So, by definition, they can be used to generate the group $SU(2^m)$ via exponentiation.

Take the antihermitian representation of all the generators defined previously:
\be\label{defT2}
\mathcal{T}^{(m)}=\{p_i\}^{(m)}\cup\{d_i\}^{(m)}.
\ee
Since the operators in $\mathcal{T}^{(m)}$ generate $SU(2^m)$, they must close to $su(2^m)$ Lie algebra; therefore
\be\label{threealgebra}
&[ p_i,p_j ]=f_{ijk} p_k ,&\nonumber\\
&[ d_i,d_j ]=g_{ijk} p_k ,&\nonumber\\
&[ p_i,d_j ]=h_{ijk} d_k ,&
\ee
with $f_{ijk}$, $g_{ijk}$, $h_{ijk}$ appropriate {\it real} structure constants.

Assume now to add {\it one single} Majorana $\gamma_{2m+1}$ to the original set $\Gamma_{2m}$.
The new Majorana set is therefore
\be
\Gamma_{2m+1}=\{\gamma_1,...,\gamma_{2m},\gamma_{2m+1}\}.
\ee
Create now the set $\{\tilde p_i\}^{(m)}$ of all the anithermitian combinations of {\it an even number} of Majoranas in $\Gamma_{2m+1}$.
These set of even operators can evidently be divided into two subsets:
\be
&&\{\tilde p_i\}=\{p_i\}\nonumber\\
&&\{\tilde{ d}_i\}=\{i \gamma_{2m+1} d_i\}.
\ee
Therefore the algebra of the operators inside $\{\tilde p_i\}^{(m)}$ descends from (\ref{threealgebra}):
\be
&[ \tilde p_i,\tilde p_j ]=f_{ijk} \tilde p_k ,&\nonumber\\
&[ \tilde{ d}_i,\tilde{ d}_j ]=g_{ijk} \tilde{ p}_k ,&\nonumber\\
&[ \tilde p_i,\tilde{ d}_j ]=h_{ijk} \tilde{ d}_k ,&
\ee
and therefore the operators of $\{\tilde p_i\}^{(m)}$ close to the Lie algebra $su(2^m)$.

It is clear that $\{\tilde p_i\}^{(m)}$ can also be obtained starting from the algebra $\mathcal{L}^{(n)}=\mathcal{L}^{(m+1)}$
obtained from the set of $2n=2m+2$ Majoranas $\Gamma_{2n}=\Gamma_{2m+2}=\{\gamma_1,...,\gamma_{2m},\gamma_{2m+1},\gamma_{2m+2}\}$,
removing all the operators that contain an arbitrarily chosen Majorana (for example $\gamma_{2m+2}$). Therefore, in the light of
the considerations of Sec.~\ref{sec:defL12}, $\{\tilde p_i\}^{(m)}$ is $\mathcal{L}^{(n)}_{1/2}$, which therefore closes to the algebra $su(2^{m})\simeq su(2^{n-1})$.

\section{Structure constant of $\mathcal{L}_{1/2}^{(n)}$}\label{app:NLTLIE}
In this appendix we will compute explicitly the Lie product:
\be
\left[p_i,p_j\right]=p_ip_j-p_jp_i, \quad p_i,p_j\in \mathcal{L}^{(n)}_{1/2}.
\ee
It is possible to see that, with this definition of the Lie product, one gets by construction:
\be\label{eq:cij}
\left[p_i,p_j\right]=c^q_{ij}p_q,\quad\text{with}\quad p_q\in\mathcal{L}^{(n)}_{1/2}.
\ee
In fact there are three possibilities:
\begin{itemize}
\item $p_i,p_j$ do not share any $\gamma_i$. That implies $c^q_{ij}=0$.
\item $p_i,p_j$ share an {\it even} number of $\gamma$-s. Also in this case $c^q_{ij}=0$, since all the non-shared couples commute.

\item $p_i,p_j$ share an {\it odd} number of $\gamma$-s.
We indicate $i_{c_1},...,i_{c_{2m+1}}$ the shared indices and by $\mu_1,...,\mu_{2(o_i-m)-1}$ and
$\beta_1,...,\beta_{2(o_j-m)-1}$ the unshared indexes of $p_i$ and $p_j$ respectively.
The numbers $2o_i$ and $2o_j$ are the orders of $p_i$ and $p_j$, which means that $o_i$ and $o_j$ count the couples of Majoranas inside  $p_i$ and $p_j$.
Then it is immediate to see that $p_ip_j$ takes the value:
\end{itemize}
\begin{widetext}
\be
p_ip_j=(-1)^m\epsilon_{\mu_1,...,\mu_{2(o_i-m)-1},i_{c_1},...,i_{c_{2m+1}}}\epsilon_{i_{c_1},...,i_{c_{2m+1}},\beta_1,...,\beta_{2(o_j-m)-1}}\mathcal{I}(o_i)\mathcal{I}(o_j)
\cdot\gamma_{\mu_1}...\gamma_{\mu_{2(o_i-m)-1}}\gamma_{\beta_1}...\gamma_{\beta_{2(o_j-m)-1}}.\nonumber\\
\ee
where we used the notation (\ref{eq:notazioneI}). The $(-1)^m$ comes from the squared values of the $m$ shared couples.
This leads to an explicit formula for $[p_i,p_j]$:
\be\label{eq:LIEproductFORMULA}
[p_i,p_j]=2(-1)^m\epsilon_{\mu_1,...,\mu_{2(o_i-m)-1},i_{c_1},...,i_{c_{2m+1}}}\epsilon_{i_{c_1},...,i_{c_{2m+1}},\beta_1,...,\beta_{2(o_j-m)-1}}\mathcal{I}(o_i)\mathcal{I}(o_j)
\cdot\gamma_{\mu_1}...\gamma_{\mu_{2(o_i-m)-1}}\gamma_{\beta_1}...\gamma_{\beta_{2(o_j-m)-1}}.\nonumber\\
\ee
\end{widetext}
It can be seen that the element $\gamma_{\mu_1}...\gamma_{\mu_{2(o_i-m)-1}}\gamma_{\beta_1}...\gamma_{\beta_{2(o_j-m)-1}}$ looks like an element of the set $\mathcal{L}^{(n)}_{1/2}$.
In fact it contains an even number of Majoranas given by $2o_q$:
\be\label{evenodd}
2o_q&=&2o_i-2m-1+2o_j-2m-1\nonumber\\
&=&2(o_i+o_j-2m-1),
\ee
and clearly if both $p_i$ and $p_j$ didn't contain a specific Majorana then also in the commutator will not contain that specific Majorana.
The only thing that should be fixed is the presence of the proper prefactor $\mathcal{I}(o_q)$ and $\epsilon_{\mu_1,...,\mu_{2(o_i-m)-1},\beta_1,...,\beta_{2(o_j-m)-1}}$
on the r.h.s of (\ref{eq:LIEproductFORMULA}). Evidently it is possible to adjust this prefactor properly to make the r.h.s. of the form $c^q_{ij}p_q$, with $p_q$
antihermitian operator in $\{p_k\}$. It is straightforward to recognize that, since $p_i$ and $p_j$ are antihermitian, then $[p_i,p_j]$ must be anithermitian too
and therefore the $c_{ij}^q$ are {\it always} real coefficients, as it should be in our case, since $\mathcal{L}^{(n)}_{1/2}$ closes to $su(2^{n-1})$ algebra and
the $p_i$ are antihermitian\cite{Cornwell, Pfeifer}.

We can fix $\mathcal{I}(o_q)$, multiplying and dividing by proper factors.
The reader can check that
\begin{itemize}
\item $o_j,o_i$ both odd, therefore $o_q$ odd: $\mathcal{I}(o_q)$ should be $+1$, which is exactly $\mathcal{I}(o_i)\mathcal{I}(o_j)$;
\item $o_j,o_i$ of opposite parity, therefore $o_q$ even: $\mathcal{I}(o_q)$ should be $+i$, which again is $\mathcal{I}(o_i)\mathcal{I}(o_j)$;
\item $o_j,o_i$ both even, therefore $o_q$ odd: $\mathcal{I}(o_q)$ should be $+1$, which is $-\mathcal{I}(o_i)\mathcal{I}(o_j)$.
\end{itemize}
\begin{widetext}
Multiplying and dividing the r.h.s of (\ref{eq:LIEproductFORMULA}) by $\mathcal{I}(o_q)$ we can identify the correct antihermitian form of $p_q$:
$$
p_q=\mathcal{I}(o_q)\epsilon_{\mu_1,...,\mu_{2(o_i-m)-1},\beta_1,...,\beta_{2(o_j-m)-1}}\gamma_{\mu_1}...\gamma_{\mu_{2(o_i-m)-1}}\gamma_{\beta_1}...\gamma_{\beta_{2(o_j-m)-1}} \in \mathcal{L}^{(n)}_{1/2}.
$$
while $c^q_{ij}$ is:
\be\label{eq:c1}
c^q_{ij}= 2\frac{\mathcal{I}(o_i)\mathcal{I}(o_j)}{\mathcal{I}(o_q)} (-1)^m\epsilon_{\mu_1,...,\mu_{2(o_i-m)+1},i_{c_1},...,i_{c_{2m+1}}}\epsilon_{i_{c_1},...,i_{c_{2m+1}},\beta_1,...,\beta_{2(o_j-m)-1}}\epsilon_{\mu_1,...,\mu_{2(o_i-m)-1},\beta_1,...,\beta_{2(o_j-m)-1}},\nonumber
\ee
\end{widetext}
which is always a real number, as we mentioned previously.
Summarizing the effect of the prefactor is:
\be\label{eq:defstrangepref}
\frac{\mathcal{I}(o_i)\mathcal{I}(o_j)}{\mathcal{I}(o_q)}=
\begin{sistema}
-1\quad\mbox{if }o_i,o_j\mbox{ are both even},\\
+1\quad\mbox{otherwise}.
\end{sistema}
\ee

Analogously one can check that
\be
c_{iq}^j=-c_{ij}^q,\quad \forall i,j,q,
\ee
which is consistent with the known properties of the $su(n)$ algebras\cite{Pfeifer}.

\section{Correlated hopping and non-linear transformations}\label{threefermions}

The correlated hopping version of the Hubbard model has been object of thorough
investigation\cite{Strack:1993fk,Ovchinnikov:1994aa,OVCHINNIKOV:1993uq,Arrachea:1994aa,Arrachea:1994ab,Airoldi:1995aa,Montorsi:1996aa,Aligia:2007aa,Montorsi:2008aa}.
The study of this system often took advantage of ``hidden'', non-linear, symmetries of the Hamiltonians, that allowed for the exact solution of the model
in some cases\cite{Arrachea:1994aa}.
So we believe it can be instructive to see how our formalism fits in this well developed context, in order to provide some insight on this general feature
of the non-linear canonical transformation. Such a discussion may also provide a hint for future studies of correlated hopping systems, which we believe may benefit from
an analysis based on the use of the hyperspin.

In Sec.~\ref{corrhoppingmain} we considered the correlated hopping model Hamiltonian
\be\label{transH}
H= -t \sum_{r,\sigma} &&\left(c^\dd_{-\sigma} \tilde c_{-\sigma}+ \tilde c^\dd_{-\sigma} c_{-\sigma}\right)\cdot\\
&&\cdot\left\{1-2(c^\dd_{\sigma}c_{\sigma}+\tilde c^\dd_{\sigma}c_{\sigma})+4c^\dd_{\sigma}c_{\sigma}\tilde c^\dd_{\sigma}\tilde c_{\sigma}\right\},\nonumber
\ee
showing that using the transformation
\be\label{definitionmapping}
R_{\chi}&=&\prod_r e^{i\chi\, (\gamma_{1,r}  \gamma_{2,r}\gamma_{3,r}\gamma_{4,r})},
\ee
with $\chi=\pi/2$, it is possible to map it into the model Hamiltonian
\be\label{free}
H^\prime=-t \sum_{r,\sigma=\pm} \left({c^\prime}^\dd_{-\sigma} {\tilde c^\prime}_{-\sigma}+ \tilde c{^\prime}^\dd_{-\sigma} {c^\prime}_{-\sigma}\right),
\ee
where we used the convention $c^\dd_\sigma=c^\dd_{r,\sigma}$, $\tilde{c}^\dd_\sigma=c^\dd_{r+1,\sigma}$.

Given this result it is possible to understand that 
if one wants some physical properties to be {\it manifestly conserved} in the final (or starting) Hamiltonians, then not all the values of $\chi$ are allowed.
Indeed, if $\chi\neq\pm\pi/2$, then a term $i\sin(\chi)({c^\prime}^\dd_{r,\sigma} {c^\prime}_{r+1,\sigma}- {c^\prime}^\dd_{r+1,\sigma} {c^\prime}_{r,\sigma})$,
appears inside the transformed Hamiltonian. This term breaks explicitly the inversion symmetry, which instead we would
(plausibly) like to have as manifestly conserved in the Hamiltonian. Therefore the only allowed values that we will consider are $\chi=\pm\pi/2$.
But since $\chi=-\pi/2$ generates the same transformation of $\chi=\pi/2$, we will analyze only this latter case in the rest of the manuscript.

It is important to remark that the non-linear transformation commutes with all the local bilinears. Therefore eventual other terms in (\ref{free}),
obtained by multiplication of local bilinears (i.e., that can be written down as the multiplication of an even number of Majoranas that belong to the same site)
do not change their form under the transformation. This is true also for the pre-factor in the correlated hopping term
\be\label{correlatedmajorana}
\begin{split}
1-2(n_\ua+\tilde{n}_\ua)+4n_\ua \tilde{n}_\ua=-\gamma_1\gamma_2\tilde\gamma_1\tilde\gamma_2,\\
1-2(n_\da+\tilde{n}_\da)+4n_\da \tilde{n}_\da=-\gamma_3\gamma_4\tilde\gamma_3\tilde\gamma_4,
\end{split}
\ee
so if the non-linear transformation is applied to (\ref{transH}), only the hopping term $c^\dd_{\sigma} \tilde c_{\sigma}+ \tilde c^\dd_{\sigma} c_{\sigma}$
is affected and generates another correlation term identical to (\ref{correlatedmajorana}).
Therefore the final coefficient of the hopping after two application of the non-linear transformation with $\chi=\pi/2$ is: 
\be
\left\{1-2(n_\sigma+\tilde{n}_\sigma)+4n_\sigma \tilde{n}_\sigma\right\}^2=1.\nonumber
\ee
This means that, starting from the free model (\ref{free}),
two subsequent applications of the non linear transformation bring the Hamiltonian back to its original form, which is not surprising since two applications
of the $\chi=\pi/2$ transformation, send $c^\dd_\sigma$ back to itself.

The Hamiltonians (\ref{free}) and (\ref{transH}) are two specific cases in the class of the correlated hopping Hamiltonians.
Assuming no translational or time reversal symmetry breaking and considering only correlation terms that commute with the local fermion density,
the most general form for the kinetic term of these Hamiltonians is\cite{Strack:1993fk,Ovchinnikov:1994aa,OVCHINNIKOV:1993uq,Arrachea:1994aa}:
\be\label{corrhopp}
H_{ch}(t_1,t_2,t_3)=\sum_{r,\sigma=\pm} &&\Bigl\{ \left(c^\dd_{-\sigma} \tilde c_{-\sigma}+ \tilde c^\dd_{-\sigma} c_{-\sigma}\right)\cdot\\
&& \cdot\left[t_1+t_2(n_\sigma+\tilde{n}_\sigma)+t_3n_\sigma\tilde{n}_\sigma\right] \Bigl\}.\nonumber
\ee
These Hamiltonians live in a three-dimensional parameter space with coordinates $(t_1,t_2,t_3)$.
The non-linear transformation creates pairwise equivalences between points of this Hamiltonian space.
A simple analysis shows that the non-linear transformation maps $H_{ch}(t_1,t_2,t_3)$ to $H_{ch}(t_1^\prime,t_2^\prime,t_3^\prime)$ as:
\be
(t_1^\prime,t_2^\prime,t_3^\prime)^T=M (t_1,t_2,t_3)^T,
\ee
with
\be\label{eq:matrixM}
M=
\begin{bmatrix}
1 & 0 & 0\\
-2 & -1 & 0\\
4 & 4 & 1
\end{bmatrix}.
\ee
%which leads to
%\be
%(t^\prime_1,t^\prime_2,t^\prime_3)=(t_1,-t_2-2t_1,t_3+4t_2+4t_1).
%\ee
It can be checked that (correctly) $M=M^{-1}$ and that $M$ has only eigenvalues equal to $\pm 1$.
Studying the The (left-)eigensystem %is
%\be\label{onebandEIGENV}
%-1 &\rightarrow& (0,1,-2),\nonumber\\
%+1 &\rightarrow& (-1,1,0),\\
%+1 &\rightarrow& (0,0,+1),\nonumber
%\ee
one understands that the correlation operator $\hat A=n_\sigma+\tilde{n}_\sigma-2n_\sigma\tilde{n}_\sigma$ is sent into minus itself by the non-linear transformation, while
the operators $\hat B=-1+n_\sigma+\tilde{n}_\sigma$ and $\hat C=n_\sigma\tilde{n}_\sigma$, are left unchanged.

The Hamiltonians (\ref{corrhopp}) can be decomposed on these $\hat A$, $\hat B$, $\hat C$ operators to make more evident these relations:
\be%\label{correlatedABC}
H_{ch}=\sum_{r,\sigma} \left\{\left(c^\dd_{-\sigma} \tilde c_{-\sigma}+ \tilde c^\dd_{-\sigma} c_{-\sigma}\right) \left(\lambda_1 \hat A+\lambda_2 \hat B+\lambda_3 \hat C\right)\right\}.\nonumber
\ee
In the three dimensional Hamiltonian space there exists a reflection symmetry with respect to the plane spanned by the eigenvactors $(t_1,t_2,t_3)=(-1,1,0)$ and $(0,0,1)$,
which means along the direction $(0,1,-2)$: Hamiltonians that have the same value of $\lambda_2$, $\lambda_3$ and $|\lambda_1|$ describe the same
physics in terms of different fermionic degrees of freedom.
This information may become extremely useful, to solve special cases or to compare how much an approximate method (numerical or theoretical) is reliable
and sensible to variations of the values of the parameters in the Hamiltonians.

The Hamiltonians with $\lambda_1=0$ are sent into themselves by the non-linear mapping. Many interesting models analyzed in the literature started from the
analysis of these special Hamiltonians, which were studied in detail in Ref.~\onlinecite{Arrachea:1994aa,Arrachea:1994ab}, although with a focus different from ours.
The interesting features of these Hamiltonians are due to the aforementioned discrete non-linear symmetry, which implies the conservation of a specific quantity.
Let us indicate the $\chi=\pi/2$ transformation with $L=R_{\chi=\pi/2}$. This operation acts on the Hamiltonian as
\be
H^\prime=L^\dd H L,
\ee
by definition. It's easier to work in terms of Majoranas, where
\be\label{eq:corrhoppNLtransf}
L&=&\prod_r e^{i\frac{\pi}{4} (\gamma_{1,r}  \gamma_{2,r}\gamma_{3,r}\gamma_{4,r})}\nonumber\\
&=&e^{i\frac{\pi}{4} \sum_{r}(\gamma_{1,r}  \gamma_{2,r}\gamma_{3,r}\gamma_{4,r})},
\ee
since the non-linear transformation acts uniformly on all the sites of the lattice.
The operator in the exponent in equation (\ref{eq:corrhoppNLtransf}) may look like the total parity operator, but it is not.
In fact, in this case, the total parity operator is given by the $P_{tot}=(-1)^N\prod_{r=1}^{N}\prod_{\alpha=1}^{4} \gamma_{\alpha,r}$, with $N$ the number of sites.
It returns $-1$ if the total number of electrons in the system is odd, otherwise it gives $+1$.
The operator $P_L=\sum_{r}\gamma_{1,r}  \gamma_{2,r}\gamma_{3,r}\gamma_{4,r}$ is instead the sum of all the local parities.
This quantity does not necessarily have to commute with the Hamiltonian (\ref{corrhopp}),
but in this specific case ($\lambda_1$=0) it does. In fact:
\be
L^\dd H L=H,
\ee
but $L^\dd=L^{-1}$, so
\be
[L^\dd, H]=[L, H]=0.
\ee
Since $L$ is not hermitian we can build the two quantities
\be\label{eq:operatoreL}
\frac{L^\dd+L}{2}&=& \cos\left(\frac{\pi}{4}\sum_r \gamma_{1,r}  \gamma_{2,r}\gamma_{3,r}\gamma_{4,r}\right),\nonumber\\
-i\frac{(L^\dd-L)}{2}&=& \sin\left(\frac{\pi}{4}\sum_r \gamma_{1,r}  \gamma_{2,r}\gamma_{3,r}\gamma_{4,r}\right),
\ee
which are therefore conserved physical quantities.
With our conventions the local parity is $-1$ for odd states and $1$ for even states;
therefore $P_L$ takes values from $-N$ to $N$, with $N$ the number of sites, and these values can change only by steps of $4$ at fixed density.
The two operators above are conserved separately, which means that $P_L$ is conserved modulo $8$, because of the $2\pi$ periodicity of the trigonometric functions.
This in turn means that the eigenstates of the Hamiltonian can be linear combinations of local states that differ by $8$ in $P_L$.
Analyzing carefully the system, one realizes that this implies a conservation {\it of the parity of the number of the doublons}, which is the
result mentioned in Sec.~\ref{corrhoppingmain}.

This result is not particularly surprising, in the light of the cited literature\cite{Arrachea:1994aa,Arrachea:1994ab,Hirsch:2002aa}, where the doublons play an
important role, also in the emergence of unconventional superconductivity.
The identification of this symmetry plays therefore an important role in the understanding of the properties of the model
and in the determination of ``special'' points in the Hamiltonian space that hide interesting physics.

If to the Hamiltonian (\ref{corrhopp}) are added a local Hubbard interaction term,
a chemical potential, magnetic fields or Boguliobov terms, all these conclusions continue to hold,
since all these terms can be written as even polynomials of local Majoranas and therefore
they commute with $L$. We stress the fact that a great deal of our conclusions regarding the Hamiltonians (\ref{corrhopp})
were already drawn in the cited literature. We just commented them in our framework and linked them to the existence of the
non-linear discrete symmetry $L=R_{\chi=\pi/2}$ in a way that is suits our discussion and is suitable of generalization.

Indeed let us consider as example a three fermion species model:
\be\label{threefermionsdef}
c^\dd =\frac{\gamma_1+i\gamma_2}{2},\quad f^\dd =\frac{\mu_1+i\mu_2}{2},\quad g^\dd =\frac{\rho_1+i\rho_2}{2}.\nonumber\\
\ee
As we have seen in Sec.~\ref{sec:section1} that in this situation there are many non-linear transformations, generated by the two set containing $15$ operators each, of the form
\be
\alpha_1&=&\gamma_1\gamma_2-i \mu_1\mu_2\rho_1\rho_2,\nonumber\\
\beta_1&=&\gamma_1\gamma_2+i \mu_1\mu_2\rho_1\rho_2,\nonumber\\
\alpha_2&=&...,
\ee
which form the two algebras $su(4)$, and by the operator
\be
\gamma_1\gamma_2\mu_1\mu_2\rho_1\rho_2,
\ee
that generates the $U(1)$ Hodge rotation. As the reader can see we omitted the trivial site index for sake of notation, since it is straightforward to understand the meaning of these
and of the following formulas.

In the objects $\alpha_i$, $\beta_i$, the bilinear part leaves the hopping term untouched, since it causes only a linear combination of the fermionic
operators. Thus the interesting part is quadrilinear component.
In order to isolate them one can combine the $\alpha_i$ and $\beta_i$ properly, obtaining:
\be
i\gamma_1\gamma_2\rho_1\mu_1,\, i\gamma_1\gamma_2\mu_1\mu_2,\, ...
\ee
In the spirit of the $n=2$ case, we are not interested in transformations that do not commute with local density terms, since they cause the appearance
of terms similar to $i\sin(\chi)(c^\dd_{r,\sigma} c_{r+1,\sigma}- c^\dd_{r+1,\sigma} c_{r,\sigma})$ into the Hamiltonian. Therefore we select the
following set,
\be
&i\gamma_1\gamma_2\mu_1\mu_2\rightarrow A,\quad i\gamma_1\gamma_2\rho_1\rho_2\rightarrow B,\quad i\mu_1\mu_2\rho_1\rho_2\rightarrow C,\nonumber&\\
&\gamma_1\gamma_2\mu_1\mu_2\rho_1\rho_2\rightarrow D,\nonumber&
\ee
as generators for the non-linear transformations of the correlated hopping model and we set the angles of the transformations to $\chi_i=\pi/2$.
The capital letters $A$, $B$, $C$, $D$ indicate the mappings that act on the correlated
hopping terms. These four mappings acts on the three kinetic terms and all commute with each other by construction, since the generators of the original
non-linear transformations commute among themselves.
When applied to $c^\dd \tilde c+\tilde c^\dd c$ one obtains the same kinetic operator multiplied by the following factors:
\be\label{relaz1}
&A&:\left[1-2\left(n_f+\tilde n_f\right)+4 n_f \tilde n_f \right],\\
&B&:\left[1-2\left(n_g+\tilde n_g\right)+4 n_g \tilde n_g \right],\nonumber\\
&C&:1,\nonumber\\
&D&:\left[1-2\left(n_f+\tilde n_f\right)+4 n_f \tilde n_f \right]\left[1-2\left(n_g+\tilde n_g\right)+4 n_g \tilde n_g \right].\nonumber
\ee
Similar results are obtained in the bands $f^\dd \tilde f+\tilde f^\dd f$, $g^\dd \tilde g+\tilde g^\dd g$, if the labels $c$, $f$, $g$, $A$, $B$, $C$ are properly exchanged.
All these factors can be computed using the Majorana representation. The relation between the transformations $A$, $B$, $C$, generated by the quadrilinears, and the $D$ one, generated
by the hexalinear, can be generalized also to models with higher number of fermions: in general, the mappings associated with a non-linear transformation of higher order
are obtained as a multiplication of the ones obtained from the quadrilinears (see Appendix~\ref{groupTR} for the demonstration).
We stress that this entire analysis is focused on the special choice of the value $\chi=\pi/2$.

It is evident that a matrix representation as (\ref{eq:matrixM}) is very useful in this analysis and to determine the
subspaces that are invariant under the four symmetry operations,as done in the $n=2$ case.
For example, consider the correlated hopping term multiplying $c^\dd \tilde c+\tilde c^\dd c$:
\be
&&\mathbf{t_1}+\mathbf{t^f_2}(n_f+\tilde n_f)+\mathbf{t^g_2}(n_g+\tilde n_g)+\mathbf{t^f_{30}}n_f\tilde n_f+\mathbf{t^g_{30}}n_g\tilde n_g\nonumber\\
&&\quad+\mathbf{t_{31}}n_gn_f+\mathbf{t_{32}}\tilde n_g\tilde n_f+\mathbf{t_{33}} n_g\tilde n_f\nonumber\\
&&\qquad +\mathbf{t_{34}}\tilde n_g n_f+\mathbf{t_{41}} n_g n_f \tilde n_f +\mathbf{t_{42}} \tilde n_g n_f \tilde n_f \nonumber\\
&&\qquad\quad +\mathbf{t_{43}} n_f n_g \tilde n_g+\mathbf{t_{44}} \tilde n_f n_g \tilde n_g+\mathbf{t_5} n_f \tilde n_f n_g \tilde n_g.\nonumber
\ee
It is easy to show that the form of the correlated hopping invariant under $A$, $B$, $C$ and therefore $D$, is
\be\label{simmetrieTHREEONE}
&&-\mathbf{t} + \mathbf{t} (n_f+\tilde n_f)+\mathbf{t} (n_g +\tilde n_g)-\mathbf{t} (n_f+\tilde n_f) (n_g +\tilde n_g)\nonumber\\
&&\quad +\mathbf{t_1}n_f \tilde n_f (n_g +\tilde n_g-1)+ \mathbf{t_2} n_g \tilde n_g (n_f +\tilde n_f-1)\nonumber\\
&&\qquad+\mathbf{t_3} n_f \tilde n_f n_g \tilde n_g.
\ee
Assuming that the symmetry between the different bands is not broken, i.e., that the correlation hopping terms are the same on each band with
an  appropriate permutation of the indices $c$, $f$ and $g$, the previous correlated hopping Hamiltonians conserve modulo 8, the quantities
\be\label{simmetrieTHREE}
&&\sum_r \gamma_1\gamma_2\mu_1\mu_2,\quad \sum_r \gamma_1\gamma_2\rho_1\rho_2,\quad \sum_r \gamma_1\gamma_2\rho_1\rho_2,\nonumber\\
&&\sum_r \gamma_1\gamma_2\mu_1\mu_2\rho_1\rho_2,
\ee
as can be shown with the same arguments used in (\ref{eq:operatoreL}).
The first three quantities are related to the symmetries $A$, $B$, $C$, and their conservation has the same meaning as in the $n=2$ fermion species case:
{\it the parity of the number of doublons on each pair of bands is conserved}. Moreover, with some straightforward
algebra and in the light of the results for the $n=2$ case, it is possible to show that the number of the $cf$, $cg$ and $fg$
doublons is also conserved if in each band $t_1=t_3=0$ or $t_1=2t$ and $t_3=-2t_2$.
%which means that the $t$, $t_1$, $t_2$ and $t_3$ are the same in each band.
If in one band these conditions are not met, then the number of two doublon species is not conserved;
if these conditions are not met in two or more bands, then no doublon species has a conserved number.
%Clearly there are many different cases and conservation rules depending upon the values of the many parameters in the different bands.

The fourth quantity in (\ref{simmetrieTHREE}), the sum of the local parities on each site, is related to the symmetry $D$.
Such symmetry is preserved also if one adds the following correlation term to the previous one:
\be
&&\mathbf{t_A} n_g n_f(1-\tilde n_f)(1-\tilde n_g)+\mathbf{t_B} \tilde n_g \tilde n_f(1- n_f)(1- n_g)\nonumber\\
&&+\mathbf{t_C} n_g \tilde n_f(1- n_f)(1-\tilde n_g)+\mathbf{t_D} \tilde n_g n_f(1- \tilde n_f)(1- n_g).\nonumber
\ee
This last term is sent into itself under $D$, but not under $A$, $B$ and $C$.
Naively one may expect that such symmetry is related with the number of triplons (coherent triple particle states analogous to the doublons) in the system.
A straightforward calculation shows that indeed the total number of triplons is conserved in the subspace where $t_A=t_B=0$ and
all the other six parameters are unconstrained. So the conservation of the number of triplons does not (in general) rely on the conservation of the number of doublons.
Anyway, the identification of the subspace that conserves the number of triplons and/or doublons, together with the identification of the non-linear symmetries of the Hamiltonian
can be a first step towards the analysis of this three fermion species model, in analogy with the $n=2$ case.
The structure of this analysis does not change increasing the number of fermion species.

A very interesting perspective is offered following the same scheme used in the $n=2$ case.
In that model, the conservation of the number of doublons was necessary to conserve a global $SU(2)$. These symmetries has been used to
solve exactly the one dimensional system\cite{Arrachea:1994aa}. One may therefore wonder if it is possible to choose proper values for the correlated hopping parameters in order
to define a global $SU(4)$ symmetry and solve analytically the problem, at least in one dimension.
This is actually not the case,  however this does not mean that it is not possible to identify a fermionic Hamiltonian with such characteristics.
We have already discussed these points in Sec.~\ref{sec.corrhopphyperspin} and we showed that it is indeed possible to build
an $SU(4)$ symmetric Hamiltonian $H_{su(4)}$ in the $n=3$ case, analogous to the $SU(2)$ symmetric one (\ref{Hn2hyper}).

In terms of the $cgf$ fermions $H_{su(4)}$ is rewritten as:
\begin{widetext}
\be
%-H_{su(4)}=t \sum_i &&\Bigl\{\left(c^\dd \tilde c+\tilde c^\dd c\right)\left[\left(n_f+\tilde n_f\right)\left(n_g+\tilde n_g\right)-\left(n_f+\tilde n_f\right)-\left(n_g+\tilde n_g\right)+1\right]\nonumber\\
%&&+\left(f^\dd \tilde f+\tilde f^\dd f\right)\left[\left(n_c+\tilde n_c\right)\left(n_g+\tilde n_g\right)-\left(n_c+\tilde n_c\right)-\left(n_g+\tilde n_g\right)+1\right]\nonumber\\
%&&+\left(g^\dd \tilde g+\tilde g^\dd g\right)\left[\left(n_c+\tilde n_c\right)\left(n_f+\tilde n_f\right)-\left(n_c+\tilde n_c\right)-\left(n_f+\tilde n_f\right)+1\right]\nonumber\\
%&&+\left(g^\dd \tilde g+\tilde g^\dd g\right)\left(f^\dd \tilde f+\tilde f^\dd f\right)\left(c^\dd \tilde c+\tilde c^\dd c\right)\nonumber\\
%&&+\frac{1}{2}\left(g^\dd \tilde g+\tilde g^\dd g\right)\left(f^\dd \tilde f^\dd+\tilde f f\right)\left(c^\dd \tilde c^\dd+\tilde c c\right)\nonumber\\
%&&+\frac{1}{2}\left(g^\dd \tilde g^\dd+\tilde g g\right)\left(f^\dd \tilde f^\dd+\tilde f f\right)\left(c^\dd \tilde c+\tilde c^\dd c\right)\nonumber\\
%&&+\frac{1}{2}\left(g^\dd \tilde g+\tilde g^\dd g\right)\left[-1+(n_c+n_f)+(\tilde n_c+\tilde n_f)-2n_c n_f-2\tilde n_c \tilde n_f\right]\nonumber\\
%&&+\frac{1}{2}\left(g^\dd \tilde g^\dd+\tilde g g\right)\left[(n_c+n_f)-(\tilde n_c+\tilde n_f)-2n_c n_f+2\tilde n_c \tilde n_f\right]
%\Bigl\}
H_{su(4)}=-t \sum_r &&\Bigl\{\left(c^\dd \tilde c+\tilde c^\dd c\right)\left[\left(n_f+\tilde n_f\right)\left(n_g+\tilde n_g\right)-\left(n_f+\tilde n_f\right)-\left(n_g+\tilde n_g\right)+1\right]\nonumber\\
&&+\left(f^\dd \tilde f+\tilde f^\dd f\right)\left[\left(n_c+\tilde n_c\right)\left(n_g+\tilde n_g\right)-\left(n_c+\tilde n_c\right)-\left(n_g+\tilde n_g\right)+1\right]\nonumber\\
&&+\left(g^\dd \tilde g+\tilde g^\dd g\right)\left[n_c \tilde n_f +\tilde n_c n_f-\frac{1}{2}(n_c+\tilde n_c)-\frac{1}{2}(n_f+\tilde n_f)+\frac{1}{2}\right]\nonumber\\
&&+\left(g^\dd \tilde g+\tilde g^\dd g\right)\left[\frac{1}{2}\left(f^\dd \tilde f^\dd+\tilde f f\right)\left(c^\dd \tilde c^\dd+\tilde c c\right)+\left(f^\dd \tilde f+\tilde f^\dd f\right)\left(c^\dd \tilde c+\tilde c^\dd c\right)\right]\nonumber\\
&&+\frac{1}{2}\left(g^\dd \tilde g^\dd+\tilde g g\right)\left[\left(f^\dd \tilde f^\dd+\tilde f f\right)\left(c^\dd \tilde c+\tilde c^\dd c\right)+(n_c+n_f)-(\tilde n_c+\tilde n_f)-2n_c n_f+2\tilde n_c \tilde n_f\right]
\Bigl\}.
\ee
\end{widetext}
Clearly this Hamiltonian does not belong to the simple correlated hopping ones analyzed previously since it contains three-body hopping operators,
but it has the $SU(4)$ symmetry analogous to the $SU(2)$ symmetry of (\ref{Hn2hyper}) with $t_3=0$.
%Also in this case a second form for this Hamiltonina, where the symmetry is not manifest, can be found following the same recipe of (\ref{secondchoice}).
Other $SU(4)$ symmetric Hamiltonians can be built, but in systems with more fermion species, enforcing projections on different subspaces via the definition of proper interactions.
The analysis will be, also in this case, very much simplified by the use of the Majorana fermion representation that
permits to identify proper Lie algebras and connect them to the fermionic representation in a straightforward way. Indeed, as we showed, building artificial (even very complciated)
Hamiltonians that obey specific symmetries becomes very easy if one uses the Majorana fermion formalism

\section{Group structure of the multi-band correlated hopping transformations}\label{groupTR}
Assume to have a set of $2n$ Majoranas $\alpha_i$ and that we have ordered them in such a way that the hopping
terms in the Hamiltonian have the form:
\be\label{salvezzatex}
-i\alpha_{2i+1}\tilde\alpha_{2i}+i\alpha_{2i}\tilde\alpha_{2i+1}.
\ee
The generators of the non-linear transformations that we consider commute with the local densities.
Therefore they have the form:
\be
i^k O_a O_b ... O_j, \quad i^k \tilde O_a \tilde O_b ... \tilde O_j,
\ee
where $k$ is the number of bilinear operators $O_i$, $O_a=\alpha_{2a}\alpha_{2a+1}$, $O_b=\alpha_{2b}\alpha_{2b+1}$, etc... and the set $\{a,b,...j\}$ is a permutation
of $k$ numbers taken from the set $\{1,...,n\}$.
The examples (\ref{corrhopp}) and (\ref{relaz1}) show well how such transformations act
on the hopping terms if applied as (\ref{definitionmapping}). Consider for example the non-linear transformation generated by $O_iO_2$. It easy to show that it transforms (\ref{salvezzatex}) into:
\be
\left(-i\alpha_{2i+1}\tilde\alpha_{2i}+i\alpha_{2i}\tilde\alpha_{2i+1}\right)\left(-O_2\tilde O_2\right).
\ee
In general, the operator $i^k O_a ... O_j$ acts as:
\begin{widetext}
\be
\left(-i\alpha_{2i+1}\tilde\alpha_{2i}+i\alpha_{2i}\tilde\alpha_{2i+1} \right) \rightarrow
 \left(-i\alpha_{2i+1}\tilde\alpha_{2i}+i\alpha_{2i}\tilde\alpha_{2i+1}\right)\left\{1+\delta_{x,i}\left[\left(-i^{2k}O_a...\hat O_x ... O_j \tilde O_a...\hat{{\tilde{O}}}_x ... \tilde O_j\right)-1\right]\right\},
\ee
\end{widetext}
where the hat means that the operator has been removed.
Take now three generators $A,B,C$ of the non-linear transformations that share $O_i$ and such that:
\be
A=O_i \mathcal{O}_A,\quad B=O_i \mathcal{O}_B,\quad C=O_i \mathcal{O}_A\mathcal{O}_B,
\ee
with $\mathcal{O}_i$ a generic multiplication of $O_a$ operators,
then it is straightforward to see that the mapping generated by $A\cdot B$ on the hopping term $-i\alpha_{2i+1}\tilde\alpha_{2i}+i\alpha_{2i}\tilde\alpha_{2i+1}$
is equal to the mapping generated by $C$ and $B\cdot A$ on the same hopping term.
Therefore the action of all mappings related to the non-linear transformations, generated by the elements that commute with the local densities,
can be obtained as a multiplication of the original mapping associated with all the non-linear transformations given by $-O_a O_b$, for all couples $a,b$.

\section*{References}
%\bibliographystyle{iopart-num}
\bibliography{Biblioteca2014}

\end{document}